\documentclass[10pt]{article}
\usepackage[margin=1in]{geometry}
\usepackage{natbib}
\usepackage{amsmath}
\usepackage{amsfonts}
\usepackage{graphicx}
\usepackage{amssymb}
\newcommand\blfootnote[1]{%
  \begingroup
  \renewcommand\thefootnote{}\footnote{#1}%
  \addtocounter{footnote}{-1}%
  \endgroup
}

\usepackage{url}
\usepackage{amsthm}
\usepackage{bm}
\usepackage[dvipsnames]{xcolor}
\usepackage{caption}
\usepackage{subcaption}

\def\E{\mathbb{E}}
\def\P{\mathbb{P}}

\newtheorem{theorem}{Theorem}[section]

\newtheorem{lemma}[theorem]{Lemma}

\newtheorem{setting}[theorem]{Setting}

\begin{document}

% Title of paper
\title{Analysis and Methods to Mitigate Effects of Under-reporting in Count Data}
% List of authors, with corresponding author marked by asterisk
\author{Jennifer Brennan$^{\ast 1}$ \and 
Marlena Bannick$^{2}$ \and
Nicholas Kassebaum$^{3}$ \and
Lauren Wilner$^{3}$ \and
Azalea Thomson$^{3}$ \and
Aleksandr Aravkin$^{4}$ \and
Peng Zheng$^{3}$}
% Author addresses

\maketitle

% Add a footnote for the corresponding author if one has been
% identified in the author list
\blfootnote{$^{1}$Paul G Allen School of Computer Science and Engineering,
$^{2}$Department of Biostatistics, 
$^{3}$Department of Health Metrics Sciences, 
$^{4}$Department of Applied Mathematics, 
University of Washington, $^\ast$\texttt{jrb@cs.washington.edu}}

\begin{abstract}
Under-reporting of count data poses a major roadblock for prediction and inference. In this paper, we focus on the Pogit model, which deconvolves the generating Poisson process from the censuring process controlling under-reporting using a generalized linear modeling framework. We highlight the limitations of the Pogit model and address them by adding constraints to the estimation framework. We also develop uncertainty quantification techniques that are robust to model mis-specification. Our approach is evaluated using synthetic data and applied to real healthcare datasets, where we treat in-patient data as `reported' counts and use held-out total injuries to validate the results. The methods make it possible to separate the Poisson process from the under-reporting process, given sufficient expert information. Codes to implement the approach are available via an open source Python package. 
\end{abstract}

\section{Introduction}
\label{sec:Intro}

Under-reporting of count data is a pervasive problem in many fields, including econometrics~\citep{winkelmann1996markov}, epidemiology~\citep{stoner2019hierarchical}, and engineering~\citep{wood2016method}. 
In population health, under-reporting of key statistics such as injuries and birth defects impedes estimation of the burden of disease~\citep{vos2020global}. 
%The observed data are the product of a true data-generating mechanism and an under-reporting mechanism, e.g., the reported rates of disease. 
The goal of statistical modeling in this area is to develop a rigorous approach to deconvolve the data-generating from the under-reporting mechanism. 
%recover the data-generating mechanism, e.g., to estimate the true rates of disease. 
However, accurately separating these two mechanisms is extremely challenging without contextual information, such as covariates related to under-reporting and those related to the true data-generating process.
% In these varied contexts, 
% researchers typically have a lot of information about 
% covariates that may be associated with the underlying rate of interest, as well as covariates that can help predict the degree of under-reporting, such as data quality variables. The main goal of statistical modeling in this area is to develop a rigorous approach to deconvolve the data-generating mechanism from the under-reporting mechanism. 

% Past work (high-level descriptions)
The predominant approach to modeling under-reported counts focuses on a two-stage model. Events are first generated according to a Poisson process with mean $\lambda$. They are then reported according to a binomial process with probability $p$, with resulting reported events having mean $\mu = \lambda p$. If covariate information is available, both $\lambda$ and $p$ can be modeled as functions of those covariates; otherwise, a mathematically convenient distribution over these parameters is assumed. When the probability of reporting is modeled using logistic regression (i.e., with a logit transform), the resulting model is called a \textit{Poisson-Logit (Pogit) model}.

% Unfortunately, separating out the two rates can be pretty unstable
Unfortunately, the deconvolution problem to separate the Poisson and bionomal processes, i.e., inferring $\lambda$ and $p$ from data that inform $\mu$, is very difficult. 
%Models for under-reported counts typically suffer from high variance when deconvolving the Poisson and binomial processes, 
% tried it, see if you like it. seems to me one can go either way, it might help to explicitly nail things down to nail the parameters but it also reads ok without. 
%\footnote{Consider rephrasing; maybe introduce $p$ and $\lambda$ earlier}. 
Since only reported counts are observed, it is challenging to determine whether these counts result from many events with a low rate of reporting (large $\lambda$, small $p$) or few events with a high rate of reporting (small $\lambda$, large $p$). This problem is exacerbated when covariates are shared between the Poisson and binomial processes or when they are highly correlated, and results in a wide range of plausible solutions. Addressing this variability, in both theory and practice, requires strong assumptions on the form of one or both latent processes.
%\sandy{In preceding paragraph, and previously, there's no need for a comma between compound objects (e.g., "between A or B."}
\paragraph{Contributions and roadmap.}
Our research makes three core contributions. First, we develop an \textbf{asymptotic covariance analysis for the Pogit model} and use it to qualitatively describe the limitations of the standard Pogit approach, which we illustrate using numerical examples. Second, we introduce new \textbf{constraints and priors to robustify the estimation process}, and we improve the  ability to deconvolve under-reporting from the true count process using rigorous  sandwich estimation to evaluate the uncertainty of model estimates. 
%\sandy{Not clear where your innovation lies...in terms of sandwich estimation. Is it a novel use of such estimation for this specific purpose? Please clarify.}
Finally, we develop an \textbf{open source implementation }of the algorithm and illustrate its successful application on a large-scale dataset that has a gold standard, validating our results. 
% very nice. 
\paragraph{Paper organization.}
Section~\ref{sec:background} describes existing models for under-reported counts. We develop an asymptotic analysis and numerical examples highlighting challenges of the deconvolution problem in  Section~\ref{sec:difficulties}. In Section~\ref{sec:methods}, we present new methods to overcome these challenges using priors and constraints as well as {novel approaches for robust uncertainty quantification}. We develop and analyze case studies using real data in Section~\ref{sec:real} and present a brief concluding discussion in Section~\ref{sec:discussion}. 
\section{{Current} models for under-reported counts}
\label{sec:background}
% Introduce the two-stage model for under-reported counts
%The problem of estimating true counts in the presence of under-reporting has been studied in multiple fields. 
%such diverse fields as econometrics, epidemiology, and highway engineering. 
The standard approach to modeling under-reported counts assumes a two-step data generating process. Let $Y_i$ represent the reported number of counts for observations $i=1,2,\ldots, n$.
%% First draw from a Poisson, then draw from a binomial
In the first step, the unobserved true number of events $Y_i^*$ is drawn according to a Poisson distribution:
\begin{align}
    Y_i^* \sim \text{Poi}(\lambda_i) \label{eqn:trueEvents}
\end{align}
In the second step, these events are filtered through a reporting process, where event $i$ has probability $p_i$ of being reported. The reported counts are modeled as a binomial random variable:
\begin{align}
    Y_i \sim \text{Binom}(Y_i^*, p_i) \label{eqn:reporting}
\end{align}
%% Turns out, this is equivalent to drawing from a poisson in the first place
This is equivalent to drawing the reported counts $Y_i$ from the Poisson distribution
\begin{align}
    Y_i \sim \text{Poi}(\lambda_i p_i).
\end{align}
%% We could use the tools of Poisson regression to estimate the product $\lambda_i p_i$ from the observations $Y_i$. The innovations in under-reported counts modeling are all in how we separate lambda_i and p_i
Estimating the reported process mean $\mu_i = \lambda_i p_i$ and its underlying process can be simply done using Poisson regression and the corresponding generalized linear model. 
%existing tools like Poisson regression, or estimating the parameters of a Poisson distribution from observations. 
However, in the under-reported counts setting, we need accurate estimates of both the true rate of events $\lambda$ and the reporting rate $p$. Previous work proposed several models to separate (deconvolve)  $p$ and $\lambda$ using observations of their product $\mu$.

% Discuss how to model $\lambda_i$ and $\p_i$; the case of whether you have covariates or not
\subsection{Models for the true rate $\lambda$ and reporting rate $p$}
The only way to deconvolve $\mu_i$ into factors $\lambda_i p_i$ is to incorporate additional assumptions about each component; otherwise, $\lambda_i = \mu_i$ and $p_i=1$ is always a valid solution. Previous work on models for under-reported counts used distributional assumptions on parameters depending on whether the observations are associated with covariates, as described below.

%% Without covariates: BB/NB model
\subsubsection{Modeling without covariates}
In the absence of covariates, a popular approach is to adopt a hierarchical model, where $\lambda$ and $p$ are latent variables drawn from underlying prior distributions. 
When a Gamma prior is placed on $\lambda$ and a Beta prior on $p$, the resulting model is called the Beta-Binomial/Negative Binomial distribution, described and analyzed in \citealp{schmittlein1985does}.
%In the Beta-Binomial / Negative Binomial Distribution model \textcolor{red}{[CITE]}\footnote{Looks like it might be a 1985 econometrics paper dealing with marketing data?}, the authors assume a Gamma prior on $\lambda$ and a Beta prior on $p$,and estimate parameters of these prior distributions using Empirical Bayes techniques. 
The specific choice of priors makes it tractable to compute posterior estimates of the individual $\lambda_i$ and $p_i$, as derived by \citealp{fader2000note} for Empirical Bayes estimation of individual $\lambda_i$ and $p_i$.
%where they are used to analyze econometrics data relating to self-reporting of port wine purchases.

%% With covariates: Pogit model
\subsubsection{Modeling with covariates}
Given covariates $x_{i, \lambda}$ that predict the true rates and covariates $x_{i, p}$ that predict reporting rates, we can model $\lambda_i$ and $p_i$ as functions of these covariates. The most popular model is the Poisson-Logistic regression, or \textit{Pogit} model, proposed by \citealp{winkelmann1993poisson}. In this model, the first step of the under-reported counts process~\eqref{eqn:trueEvents} is modeled according to standard Poisson regression with coefficients $\theta_\lambda$:
\begin{align}
    Y_i^* \sim \text{Poi}\left(\exp\left({x_{i,\lambda}^\top\theta_\lambda}\right)\right).
\end{align}
The second step~\eqref{eqn:reporting} is modeled according to logistic regression with coefficients $\theta_p$:
\begin{align}
    Y_i &\sim \text{Binom}\left( Y_i^*, \frac{\exp(x_{i, p}^\top \theta_p)}{1 + \exp(x_{i, p}^\top \theta_p)} \right)\\
    &=: \text{Binom}\left( Y_i^*, \text{expit}\left( x_{i, p}^\top \theta_p \right) \right).
\end{align}
The generating distribution for $Y_i$ can now be written as the Poisson distribution
\begin{align}
\label{eq:params}
    Y_i \sim \text{Poi}\left( \exp\left({x_{i,\lambda}^\top\theta_\lambda}\right) \text{expit}\left( x_{i, p}^\top \theta_p \right)\right).
\end{align}
The Pogit model has been used to estimate worker absenteeism in econometrics \citep{winkelmann1996markov}, tuberculosis incidence in epidemiology \citep{stoner2019hierarchical}, and traffic accidents in highway engineering \citep{wood2016method}. 

\subsection{Parameter Estimation}
% Discuss ways to fit the model
Estimating the parameters of the Pogit model remains a deconvolution problem of $\lambda$ and $\mu$ processes: a high observed count may be due to either a high underlying rate or a high reporting rate. Previous work addressed identifiability conditions for the Pogit model as well as several ways to include side information to improve the parameter estimates.
% De-emphasize frequentist vs Bayesian
%The parameters of the Pogit model can be estimated in either the maximum likelihood or the Bayesian framework. Previous work in both of these settings has identified challenges with high variance when separating $p$ and $\lambda$.

%% MLE
\subsubsection{Conditions for parameter identifiability}\label{subsec:identifiability}
%%% Discuss identifiability paper: can't use all covariates for both processes; there are two equivalent parameterizations; discussion about "weak" identifiability (qualitative)
A complete treatment of the difficulties in separating $p$ and $\lambda$ under the maximum likelihood framework is given by \cite{papadopoulos2012identification}, who observed  two distinct Pogit model parameterizations that lead to the same conditional law $\P(Y | x)$:
\begin{align}
\label{eq:idenIssue}
    \mu_i := \exp(x_{i,\lambda}^\top \theta_\lambda) \frac{\exp(x_{i,p}^\top\theta_p)}{1+\exp(x_{i,p}^\top\theta_p)}
    = \exp(x_{i,\lambda}^\top \theta_\lambda + x_{i,p}^\top\theta_p) \frac{\exp(-x_{i,p}^\top\theta_p)}{1+\exp(-x_{i,p}^\top\theta_p)}
    =: \mu_i^a.
\end{align}
The authors show that identifiability can be regained either by knowing the sign of some nonzero element of $\theta_p$ a priori or by restricting some covariates to $x_{i,p}$, excluding them from $x_{i, \lambda}$. In earlier work \citep{papadopoulos2008identification}, the authors discussed problems that could arise if the restricted covariate is nearly colinear with the remaining covariates; in this case, they illustrated the resulting near-unidentifiability using employment data from the German Socio-Economic Panel (\citealp{wagner1993english}), which was previously analyzed in \citealp{winkelmann2008econometric}.

The identifiability problem~\eqref{eq:idenIssue} shows how the inherent ambiguity in deconvolving a product into individual terms directly translates to ambiguity in the Pogit model. 
In this work, we reduce this ambiguity by incorporating additional information in the parameter estimation, i.e., using constraints and regularization, to better resolve $p$ and $\lambda$ (and their associated models) in the maximum likelihood framework.
%To resolve this ambiguity, we develop an approach to incorporate additional information using constraints to resolve $p$ and $\lambda$ via a maximum likelihood approach. 

% Methods for reducing variance
\subsubsection{Variance reduction methods}
Correlation among covariates for $\lambda$ and $p$ increases the risk of unidentifiability in the model 
and can manifest as high variance, as noted by ~\cite{papadopoulos2012identification}. 
 Another source of variance in the Pogit model is model misspecification, which can occur if there is overdispersion in the Poisson process. Several techniques have been developed to address these sources of variance. 

One technique adds constraints or priors to the model, incorporating side information to reduce variance. This approach, used by \cite{stoner2019hierarchical}, applies the Pogit model to the problem of estimating tuberculosis incidence in regions of Brazil using a Bayesian formulation. Here, the side information is a prior on the aggregate rate of tuberculosis reported across all regions, elicited from WHO estimates. The authors emphasize the strong dependence of the fitted model on this prior.
%, which points to its importance as a regularizer.
Another type of side information useful for reducing variance is the presence of fully reported observations for which the reporting rate is one. This type of regularization {induces the function $p$ to pass through certain points} and is used in the analyses of \citealp{stamey2006bayesian} and \citealp{dvorzak2016sparse}.

A second technique addresses model misspecification by increasing the Pogit model's flexibility. Overdispersion of the true observations could be addressed by replacing the Poisson model with a negative binomial, although to our knowledge this has not been done previously.
In the Bayesian setting, \cite{stoner2019hierarchical} address overdispersion by including additional Gaussian noise in the relationship between $\lambda$ and $p$ and their covariates.

%% Bayesian methods (the TB paper)
%\subsubsection{Bayesian Methods}
%%% Discuss sensitivity to priors
% \cite{stoner2019hierarchical} apply the Pogit model to the problem of estimating tuberculosis incidence in the regions of Brazil. 
% They apply Bayesian hierarchical methods to model the incidence and reporting of tuberculosis as a function of regional covariates, with spatial correlation. 
% To avoid the identifiability issues presented in \cite{papadopoulos2012identification}, they restrict each covariate to only enter the model in either $p$ or $\lambda$, but never both. 
% Since an intercept term was present in the models for both $p$ and $\lambda$, the paper emphasized the importance of a strong prior on at least one intercept term. 
% This was accomplished for the tuberculosis data by eliciting a strong prior on the average reporting rate of tuberculosis in Brazil, which was taken from WHO estimates. 
% If this type of strong prior is readily accessible to practitioners, then it provides valuable information to disentangle the true counts from the reporting rate. However, this type of prior information is not always available.

\section{Characterizing the difficulties of $p$, $\lambda$ deconvolution}
\label{sec:difficulties}
%Previous work on the Pogit model discussed its high variance in separating $p$ and $\lambda$, especially in the case where $p$ and $\lambda$ depend on shared or closely correlated covariates. 
With the exception of \cite{papadopoulos2012identification}, no work has analyzed the shortcomings of the Pogit model from a theoretical perspective. In Section~\ref{sec:lowerBound}, we derive an asymptotic lower bound for the variance of the maximum likelihood estimate of the Pogit parameters under a simplified setting, where $p$ and $\lambda$ each depend on a single covariate. This analysis reveals a fundamental difficulty: the variance of $\theta_p$ grows with $\theta_p^2$, making it difficult to identify $p$, and hence $\lambda$, in a setting with moderate or large $\theta_p$. We test this intuition using numerical simulations in  Section~\ref{sec:sim}, where we present a simple setting that nonetheless makes it impossible to infer even the sign of $\theta_p$ for any value of $\theta_p$.

\subsection{Theoretical analysis in the two-covariate setting}
\label{sec:lowerBound}
To characterize the behavior of the estimated Pogit model, we analyze a simple version of it. In our setting, $p$ and $\lambda$ are each determined by a single covariate
%% Equation for Y
\begin{align}
    p_i &= \frac{\exp(x_{p, i} \theta_p)}{1+\exp(x_{p,i} \theta_p)}\\
    \lambda_i &= \exp(x_{\lambda, i}\theta_\lambda)
\end{align}
so that
\begin{equation}
    \label{eqn:simplePogit}
    Y_i\sim\text{Poi}\left( 
    \exp(x_{\lambda, i}\theta_\lambda) 
    \frac{\exp(x_{p, i} \theta_p)}{1+\exp(x_{p,i} \theta_p)}
    \right). 
\end{equation}
We are interested in the performance of the estimators of parameters $\bm\theta = [\theta_p, \theta_\lambda]$, measured by the mean squared error
\begin{align}
    \text{MSE}(\hat\theta) &:= \mathbb{E}_{Y, X}\left[ (\hat\theta - \theta)^2 \right].
\end{align}
%% Conditions on x
% Since the MSE is an expectation taken over both the observations $Y$ and the covariates $X$, we must specify a distribution over $X$ in addition to a model for $Y$. We choose 
% \begin{equation}
% \label{eq:setting}
% \begin{aligned}
%     x_\lambda &\sim \mathcal{N}(0, 1)\\
%     x_p &\sim \mathcal{N}(0,1) 
% \end{aligned}
% \end{equation}
For $i=1,2,\ldots, n$, let covariates $x_{p, i}$ and $x_{\lambda, i}$ be drawn independently according to
\begin{equation}
    \label{eq:setting}
\begin{aligned}
    x_{\lambda, i} &\sim \mathcal{N}(\mu_\lambda, \sigma_\lambda^2)\\
    %x_{p, i} &\sim \mathcal{D}_p \qquad \text{s.t.}~\mathbb{P}_{\mathcal{D}_p}(x) =\mathbb{P}_{\mathcal{D}_p}(-x) 
    x_{p, i} &\sim \mathcal{N}(0, \sigma_p^2)
\end{aligned}
\end{equation}
% and let $Y_i$ be drawn according to
% \begin{align}
%     Y_i &\sim \text{Poi}\left( e^{x_{\lambda, i}\theta_\lambda} \frac{\exp\left( x_{p, i}\theta_p \right)}{1 + \exp\left( x_{p, i}\theta_p \right)} \right)
% \end{align}
% Since we are fitting our model with MLE, which is asymptotically unbiased (as long as we have identifiability! which is true in this specific model), it remains to understand the asymptotic variance.
To analyze the behavior of the maximum likelihood estimates of $\theta_p$ and $\theta_\lambda$, we assume as a further technical condition that $\theta_\lambda, \theta_p \in [C_l, C_u]$ for some constants $C_l, C_u \in \mathbb{R}$. 
These assumptions help us prove regularity conditions about the maximum likelihood estimator, but in practice they can be chosen as sufficiently large 
to be inactive at the solution.

First, we provide preliminary results about the Pogit model. We did not find these results in the literature and include them here for completeness, with proofs in Appendix~\ref{app:lowerBound}. 

Recall that the Fisher information matrix is defined by 
\begin{align}
    \mathcal{I}(\theta) = \E\left[ (\nabla_\theta
    \log f(X,Y;\theta))(\nabla_\theta\log f(X,Y;\theta))^\top \Big| \theta \right],
\end{align}
where $f$ is the probability density function of the data $X$ and $Y$ given parameters $\theta$. We now state the following lemma, which contains the regularity conditions required (1) to show that the MLE is asymptotically normally distributed, and (2) for the Cram\'er-Rao lower bound to hold:
%\sandy{Verify tech accuracy of edit. As written, it was difficult for me to parse.}

% \begin{lemma}\label{lemma:conditions}
% Let $\{x_{\lambda, i}, x_{p,i}, Y_i \}_{i=1}^n$ be drawn as in~\eqref{eq:setting},
% and consider the likelihood $\ell_{\bm\theta}$
% corresponding to~\eqref{eqn:simplePogit}.
% %Setting \ref{set:twoCovPogit}. 
% Then the following two regularity conditions are satisfied:
% \begin{enumerate}
%     \item The Fisher information is always defined, i.e. 
%     \[
%     \nabla_{\bm\theta} \log \ell_{\bm\theta}(Y, X)
%     \]
%     exists and is finite for all $X$, $Y$ such that the likelihood $\ell_{\bm\theta}(Y, X)$ is positive.
%     \item The operations of integration with respect to $x$ and $y$, and differentiation with respect to $\bm\theta$, can be interchanged in the expectation of $\bm{\hat\theta}$:
%     \[
%     \nabla_{\bm\theta}\left[\int \int \bm{\hat\theta} \ell_{\bm\theta}(x, y) dxdy \right]
%     =
%     \int \int \bm{\hat\theta} \left[ \nabla_{\bm\theta} \ell_{\bm\theta}(x, y) \right] dx dy.
%     \]
% \end{enumerate}
% \end{lemma}
%Both regularity conditions are required for the Cram\'er-Rao lower bound, %and the second condition is required to rewrite the Fisher information matrix in a convenient form.
%These conditions are required to
%and to show the MLE is asymptotically unbiased, as stated formally in the following lemma. 

\begin{lemma}\label{lemma:mleNormConditions}
Let $\{x_{\lambda, i}, x_{p,i}, Y_i \}_{i=1}^n$ be drawn as in~\eqref{eq:setting}. The following regularity conditions hold.

\begin{enumerate}
    \item $\theta_0$ is \textit{identified} such that if $\theta\neq \theta_0$ and $\theta\in\Theta$, then $\ell(x,y|\theta) \neq \ell(x,y|\theta_0)$ with respect to the dominating measure $\mu$.
    \item $\mathbf{\theta_0}$ lies in the interior of $\Theta$, which is assumed to be a compact subset of $\mathbb{R}^2$.
    \item $\log \ell(x, y|\theta)$ is continuously differentiable at each $\theta\in\Theta$ for all $x, y \in \mathcal{X}\times \mathcal{Y}$ (a.e. will suffice).
    \item {$|\log \ell(x, y | \theta)| \leq d(x, y)$ for all $\theta\in\Theta$ and $\E_{\theta_0}[d(X,Y)] < \infty$.}
    \item $\ell(x, y | \theta)$ is twice continuously differentiable, and $\ell(x, y | \theta) > 0$ in a neighborhood, $\mathcal{N}$, of $\theta_0$.
    \item $||\tfrac{\partial \ell(x, y | \theta)}{\partial \theta}|| \leq e(x, y)$ for all $\theta\in \mathcal{N}$ and $\int e(x, y) d\nu(x, y) < \infty$.
    \item Defining the score vector
    \[
    \psi(x,y | \theta) = (\partial \log \ell(x,y|\theta) / \partial \theta_1, \ldots, \partial \log \ell(x,y | \theta) / \partial \theta_k)',
    \]
    we have that
    $I(\theta_0) = \E_{\theta_0} [\psi(X,Y|\theta_0)\psi(X,Y|\theta_0)']$ exists and is non-singular.
    \item {$|| \tfrac{\partial^2 \log \ell(x,y|\theta)}{\partial \theta \partial \theta'}|| \leq f(x,y)$ for all $\theta\in \mathcal{N}$ and $\E_{\theta_0} [f(X,Y)] < \infty$.}
    \item $|| \tfrac{\partial^2 \ell(x,y|\theta)}{\partial \theta \partial \theta'}|| \leq g(x,y)$ for all $\theta\in \mathcal{N}$ and $\int g(x,y) d\nu(x,y) < \infty$.
\end{enumerate}

And the maximum likelihood estimate of $\mathbf{\theta}$ is distributed according to
\begin{align}
    \mathbf{\hat\theta}_{MLE} \sim \mathcal{N}\left(\mathbf{\theta}, \mathcal{I}(\mathbf{\theta})^{-1}\right).
\end{align}
\end{lemma}

% In Appendix~\ref{app:lowerBound}, we show 
% that the model of Eqn \eqref{eqn:simplePogit} satisfies regularity conditions that ensure the MLE is asymptotically unbiased.
% % We can do this (asymptotically) using the Cramer Rao lower bound (note, we satisfy the regularity conditions)
% We can therefore  focus the analysis on the asymptotic covariance matrix of the MLE. Under mild conditions, which we show~\eqref{eqn:simplePogit} satisfies, the MLE converges in distribution to 
% \begin{align}
%     \hat\theta_{MLE} \sim \mathcal{N}\left( \theta, \mathcal{I}(\theta)^{-1} \right)
% \end{align}
 %In addition to describing the asymptotic coviariance of the maximum likelihood estimates of the parameter vector $\theta$, 
The inverse of the Fisher Information Matrix provides a lower bound on the variance of any unbiased estimator via the Cram\'er Rao lower bound; a lower bound on the Fisher Information Matrix therefore provides insight into the fundamental difficulty of the estimation task. Our main result is given below. 

\begin{theorem}\label{thm:fullGenerality}
Let $\{x_{\lambda, i}, x_{p,i}, Y_i \}_{i=1}^n$ be drawn as in~\eqref{eq:setting}. %Setting \ref{set:twoCovPogit}.
Let $\bm{\hat\theta}$ be any unbiased estimator of $\bm{\theta}=[\theta_\lambda, \theta_p]$. Then, the covariance of $\bm{\hat\theta}$ is lower bounded by
\begin{align}
\label{eq:boundresult}
    \text{Cov}(\bm{\hat\theta}) 
    &\succeq
    \frac{1}{n\E[\lambda]}
    \begin{bmatrix}
    \frac{1}{\E[p]\left(\left(\mu_\lambda + \sigma_\lambda^2\theta_\lambda\right)^2 + \sigma_\lambda^2\right)} & 0 \\
     0 & 2\theta_p^2
    \end{bmatrix}.
\end{align}
\end{theorem}

%how hard the estimation task is for a variety of estimation procedures.

To gain a more intuitive understanding of the result in Theorem~\ref{thm:fullGenerality}, we take $\sigma^2_\lambda =\sigma_p^2=1$ and $\mu_\lambda = 0$, in which case
%~\eqref{eq:boundresult} simplifies to 
%\footnote{Probably make this a theorem statement or something}
% \begin{align}
% \text{Cov}(\bm{\hat\theta}) 
% &\succeq 
% \frac{1}{n\mathbb{E}[\lambda]}
% \begin{bmatrix}
% \frac{2}{\left( \theta_\lambda^2 + 1\right)}
% & 0\\
% 0 &
% 2 \theta_p^2 
% \end{bmatrix}
% \end{align}
the asymptotic variance for $\hat{\theta_\lambda}_{MLE}$ and $\hat{\theta_p}_{MLE}$ can be lower-bounded by
\begin{align}
    \text{Var}(\hat{\theta_\lambda}_{MLE}) &\geq \frac{2}{n\E[\lambda]\left( \theta_\lambda^2 + 1\right)}\\
    \text{Var}(\hat{\theta_p}_{MLE}) &\geq \frac{2}{n\E[\lambda]}\theta_p^2
\end{align}

% Interpret the covariance... Also what does it say about p and lambda?
The standard deviation in estimating $\theta_p$ grows at least proportionally to the value of $\theta_p$ itself; for a fixed value of $n$, there will be at least a constant probability of incorrectly estimating the sign of $\theta_p$, which is equivalent to incorrectly determining whether $p$ is an increasing or decreasing function of $x_p$. 
%In many global health contexts, estimating $p$ is not as important as estimating $\lambda$, but 
In applications where estimating $p$ is important, additional information and constraints {\it must} be used to decrease the variance of $\hat{\theta_p}_{MLE}$.
%\footnote{at the expense of additional bias, of course} 

%\textcolor{red}{TODO: Some comment on estimating $\theta_\lambda$, like how the variance gets better if $\lambda$ and $p$ are both large (as you would expect)?}

%\textcolor{red}{TODO: FYI, we have E[lambda]E[p] = E[mu]}

\subsection{Simulation Study}
\label{sec:sim}

We perform a simulation study to illustrate the high variance of $p$ and $\lambda$ in this simple setting. For our simulations, we generate $n=50$ covariates $x_\lambda \sim \mathcal{N}(0,1)$ and $x_p \sim \mathcal{N}(0,1)$. 

\begin{figure}[h!]
    \centering
    \includegraphics[width=0.3\textwidth]{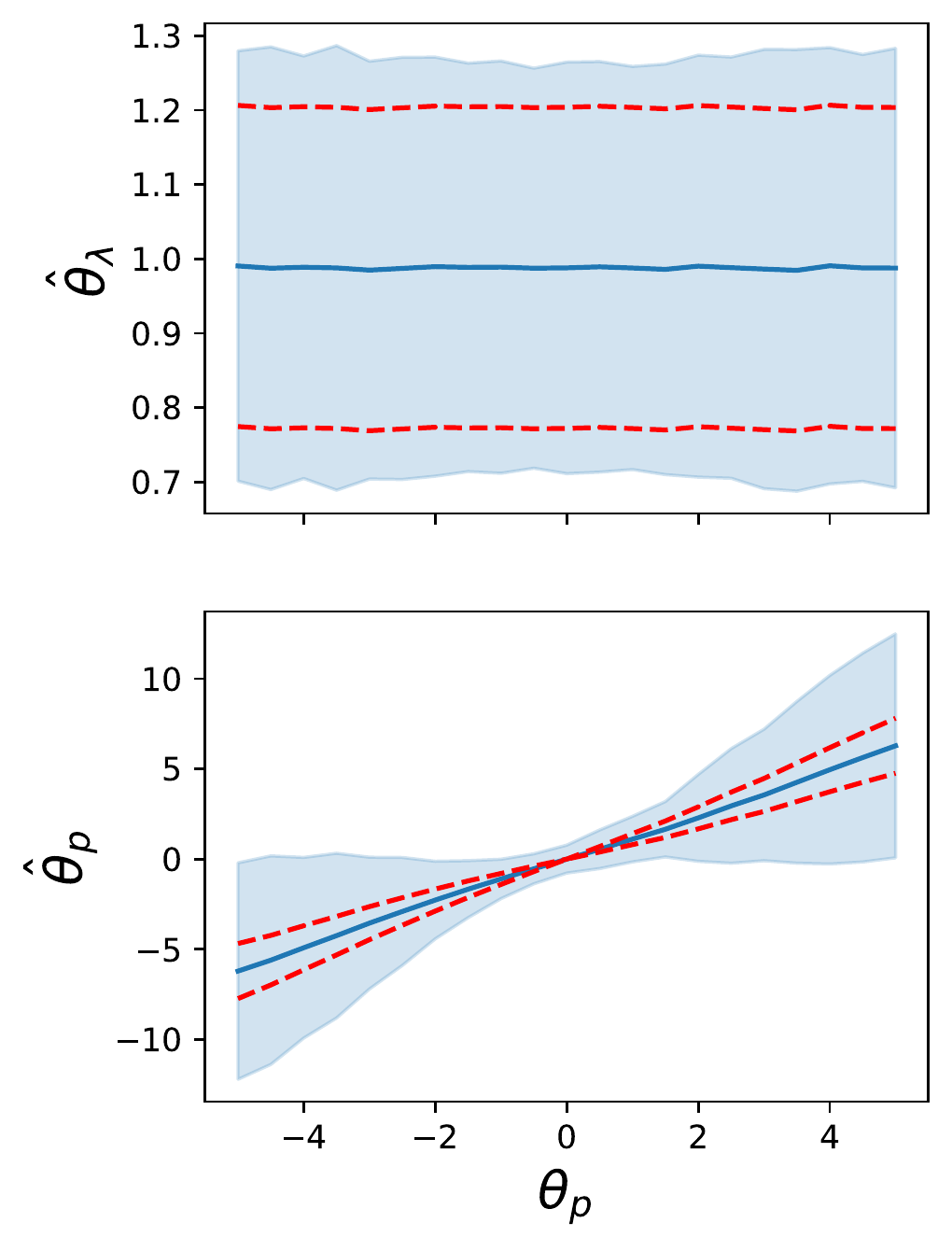}
    \qquad
    \includegraphics[width=0.3\textwidth]{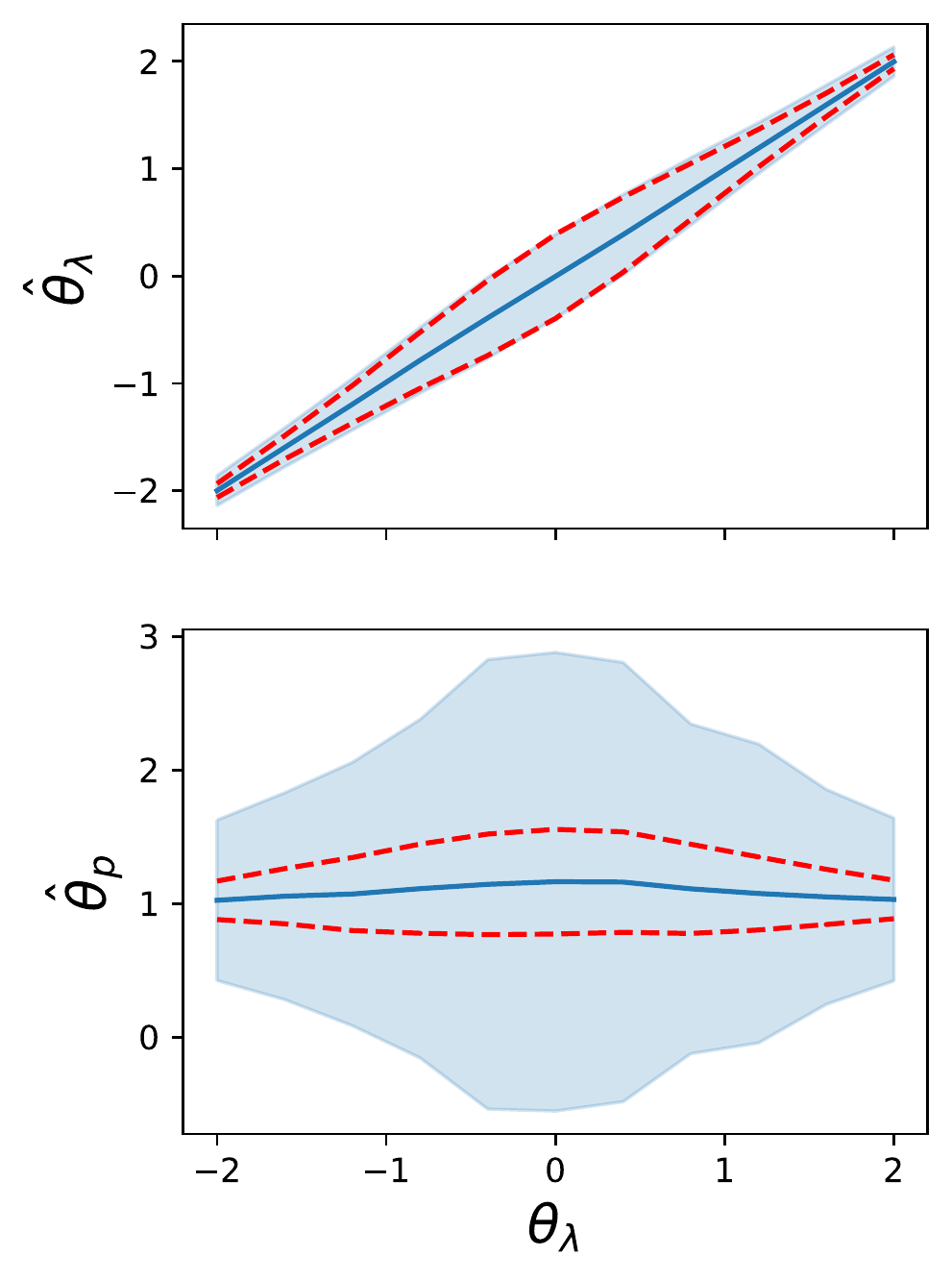}
    \caption{Simulation studies illustrating the dependence of $\text{Var}(\hat\theta_{p,MLE})$ and $\text{Var}(\hat\theta_{\lambda,MLE})$ on $\theta_p$ and $\theta_\lambda$ for $n=50$ data points. The left panels vary $\theta_p$ with fixed $\theta_\lambda=1$, while the right panels vary $\theta_\lambda$ with fixed $\theta_p=1$.
    Solid blue lines show the average parameter estimate over $5000$ trials; the blue shaded region is the 95\% confidence interval based on the empirical standard deviation. The red dashed lines show our theoretical lower bound on the 95\% confidence interval.}
    \label{fig:simpleSim}
\end{figure}

For the first simulation, shown in Figure \ref{fig:simpleSim} (left), we fix $\theta_\lambda = 1$ and vary $\theta_p$ from $-5$ to $5$, then we report the 95\% confidence interval based on the empirical standard deviation of $\hat\theta_p$ and $\hat\theta_\lambda$ over $5000$ replicates. We compare this to our theoretical lower bound on the confidence interval for each parameter. Our experiments confirm the theoretical bound in this setting: the standard deviation in $\hat\theta_\lambda$ is independent of $\theta_p$, while the standard deviation of $\hat\theta_p$ grows linearly with the parameter $\theta_p$, highlighting the difficulty of estimating $p$ in this setting. We also see that $\theta_p=0$ is always within the 95\% confidence interval, indicating that no matter how large $\theta_p$ truly is, we cannot determine whether $p$ increases or decreases with $x$ in this setting.

We perform a second simulation with $\theta_p=1$ fixed while $\theta_\lambda$ varies from $-2$ to $2$, shown in Figure \ref{fig:simpleSim} (right). The results confirm our interpretation of the lower bound; the standard deviation in both $\hat\theta_\lambda$ and $\hat\theta_p$ decreases as the magnitude of $\theta_\lambda$ increases.

These simulations illustrate that the estimated parameters can suffer from high variance;
in particular, we can never reject the hypothesis that $\theta_p = 0$ for any value of $\theta_p.$ In practical settings, we need additional information to reduce the variance in the model and successfully deconvolve $p$ and $\lambda$. We next discuss a framework for incorporating priors and constraints as a powerful way to bring expert knowledge to bear on specific problems. 

\section{Incorporating prior knowledge into model building}
\label{sec:methods}

We now discuss how we use prior knowledge to deconvolve the under-reporting and data-generating mechanisms while maintaining model identifiability (Section  ~\ref{sec:covariates}) and how we apply the sandwich estimation procedure to quantify the resulting model's uncertainty (Section~\ref{sec:UQ}).

%estimate uncertainty. Section  ~\ref{sec:covariates}  describes techniques to flexibly model mechanisms for both under-reporting and data-generating using covariates while maintaining model identifiability. Section~\ref{sec:UQ} presents the sandwich estimation procedure we use to quantify uncertainty for the resulting model. 

% (1) a general mechanism to flexibly model both the under-reporting and the data-generating mechanisms with covariates, (2) identifiability of the model, and (3) uncertainty quantification. In Section~\ref{sec:covariates}, we discuss a spline model that allows us to infer a more detailed relationship between covariates and key rates. 

\subsection{Covariate Specifications}
\label{sec:covariates}

To deconvolve the under-reporting mechanism (parametrized by $p$) from the data-generating mechanism (parametrized by $\lambda$), we model both as functions of covariates. These functions consist of a linear predictor and a link function that transforms the linear predictor to the desired space, as in generalized linear models. For example, to model the parameter $\lambda$ using a vector of covariates $\bm x$, a vector of regression coefficients $\bm \theta$, and link function $g$:
\begin{align*}
    \lambda := g^{-1}(\bm x^\top \bm \theta).
\end{align*}

\paragraph{Linear predictors.} For $\bm x^\top \bm \theta$, we can choose simple covariate specifications, like including a continuous predictor as a single linear term. Alternatively, nonparametric regression techniques (such as basis splines~\citep{de1978practical}) let us flexibly parametrize the relationship between a covariate and $\lambda$ or $p$. These spline specifications, which include the degree of the spline and the number and location of knots, are embedded in the linear predictor $\bm x^\top \bm \theta$.

To encode knowledge about the shape of the relationship among covariates, we can use linear inequality constraints on the regression coefficients $\bm \theta$. For example, such constraints could force the regression coefficient to be positive, which would enforce an increasing relationship between the true rate of reporting and a given covariate. Further, very general linear constraints are particularly useful for working with basis splines. The second derivative of a basis spline can be represented as a linear function of its basis elements, so linear constraints of the regression coefficients let us constrain the second derivative to be positive or negative.

Finally, rather than constraining the regression coefficients $\bm \theta$ of the linear predictor, we can include a quadratic regularizer, commonly known as a Gaussian prior or `ridge' regression penalty~\citep{hoerl1970ridge}.
Trading off bias for variance in the parameter estimation, we can 
avoid over-fitting the data, and we can incorporate prior beliefs in a quantitative way.

% \paragraph{Linear constraints.} Linear constraints are a more rigid form of regularization compared to using a penalty,
% and are particularly useful when working with splines. For example, we can include monotonic constraints of the spline to 

\paragraph{Link functions.} In the Pogit model, we use different link functions $g$ for $\lambda$ and $p$.
Specifically, we use the logit function for $p$ and the log function for $\lambda$.
%The differences between the two link functions help with the identification problem, compared to a pure Poisson model.
% ^ I think we would need to explain the "pure Poisson model" if we wanted to include this
W can also use other functional forms,as needed. For example, if we understand from prior knowledge that the under-reporting rate is between $a$ and $b$, the inverse link function
\[
l_p(x) = a + \frac{b - a}{1 + \exp(-x)}
\]
for 
parameter $p$ captures this information with no need for more complex constraints.

% to encode expert knowledge about these relationships. 
% regularizers and constraints 

% We use basis splines  to flexibly parametrize the relationship between the covariates and the parameters $(\lambda, p)$. B-splines represent nonlinear functions as linear combinations of spline basis elements. The result is a set of piecewise polynomials.

% \paragraph{Link functions.}

% For polynomial with degree $p$ and number of knots $k$, we recursively generate $p+k$ basis 
% elements $s^p_{j}$.

% With this basis, we represent a nonlinear curve as the linear combination of the basis elements, 
% with coefficients $\theta \in \mathbb{R}^{p+k}$: 
% \begin{equation}
% \label{eq:spline}
% f(t) = \sum_{j=1}^{p+k} \theta_j^p s_j^p (t).
% \end{equation}
% An explicit representation of~\eqref{eq:spline} is obtained by building a design matrix $X$.
% The $j$th column of $X$ 
% is given by evaluating the basis elements on domain points where we have observations:
% \begin{equation}
% \label{eq:splineMatrix}
% X_{\cdot, j} = \begin{bmatrix} s_j^p(t_0) \\ \vdots \\ s_j^p(t_k)\end{bmatrix}.
% \end{equation}
% These columns can be used to parametrize $\lambda$ or $p$ as in~\eqref{eq:params}.

% \subsection{Regularizers and Constraints}
% \label{sec:constraints}

% We provide three types of regularization: selection of the link function, quadratic regularizers (Gaussian priors), and linear inequality constraints.

% \paragraph{Quadratic regularizers.} 

\paragraph{Example.} To show the impact of the innovations on the pogit model, we use a simple synthetic example, where $p$ and $\lambda$ are taken to be simple nonlinear functions 
\[
\lambda = 15 + \exp\left(\cos(2\pi x_0)\right), \quad p = \text{expit}\left(\sin(2\pi x_1) \right),
\]
with $x_0$ and $x_1$ simulated as independent uniform random variables. The Pogit model is fitted using only reported data. 
Spline specifications for $p$ and $\lambda$ are used to capture the nonlinear relationships. Figure~{fig:constraintComparisons} shows the results for predicted $p, \lambda$, and $\mu = p\lambda$ across 100 realizations of the experiment. Its first column presents results for the unconstrained spline Pogit approach; though the $\mu$ fit is correct (third row), resolving $p$ and $\lambda$ is far more difficult. In each column thereafter, we show the impact of the three techniques described above. The table's second column shows results for the modified link function that bounds $p$ between $0.2$ and $0.8$ through its representation. The third column shows results for using quadratic regularization pulling $p$ to $0.5$. 
Finally, the fourth column presents imposing convexity constraints on $p$ (as a function of $x_1$) and $\lambda$ (as a function of $x_0$). 
All three techniques improve resolution of $p$ and $\lambda$, with quadratic regularization helping the most: it provides specific (strong) information about the value of $p$ rather than general (weaker) information about the bounds on $p$ or the nature of the relationship between $p$ and $x_1$ or $\lambda$ and $x_0$. All approaches are comparable in their recovery of $\mu$, which underscores the fact that the $\mu$ fit alone cannot differentiate successful resolution of $p$ and $\lambda$ (e.g., as in column 3) and failure to resolve these parameters (as in the unconstrained results of column 1).

% \textcolor{red}{For each novelty, what do you think of making a simple example that shows the failure of `naive' recovery, and success of modified link function, regularizer, or constraint to get it? I know we have very interesting case studies but maybe super simple plots for each subsection showing the effect could make it a more exciting read. We can also combine all of these into a figur for page 3, right around the contributions.  }
% Yeah, this sounds like a good idea. We could do the simple two-covariate setting, so the plots are easy to read as well.
% Honestly, the shape constraints didn't do much here. I have an idea that if we dial down the number of observations, it might start getting confused about the shape, and maybe that will help.
\begin{figure}[h!]
    \centering
    \includegraphics[width=\textwidth]{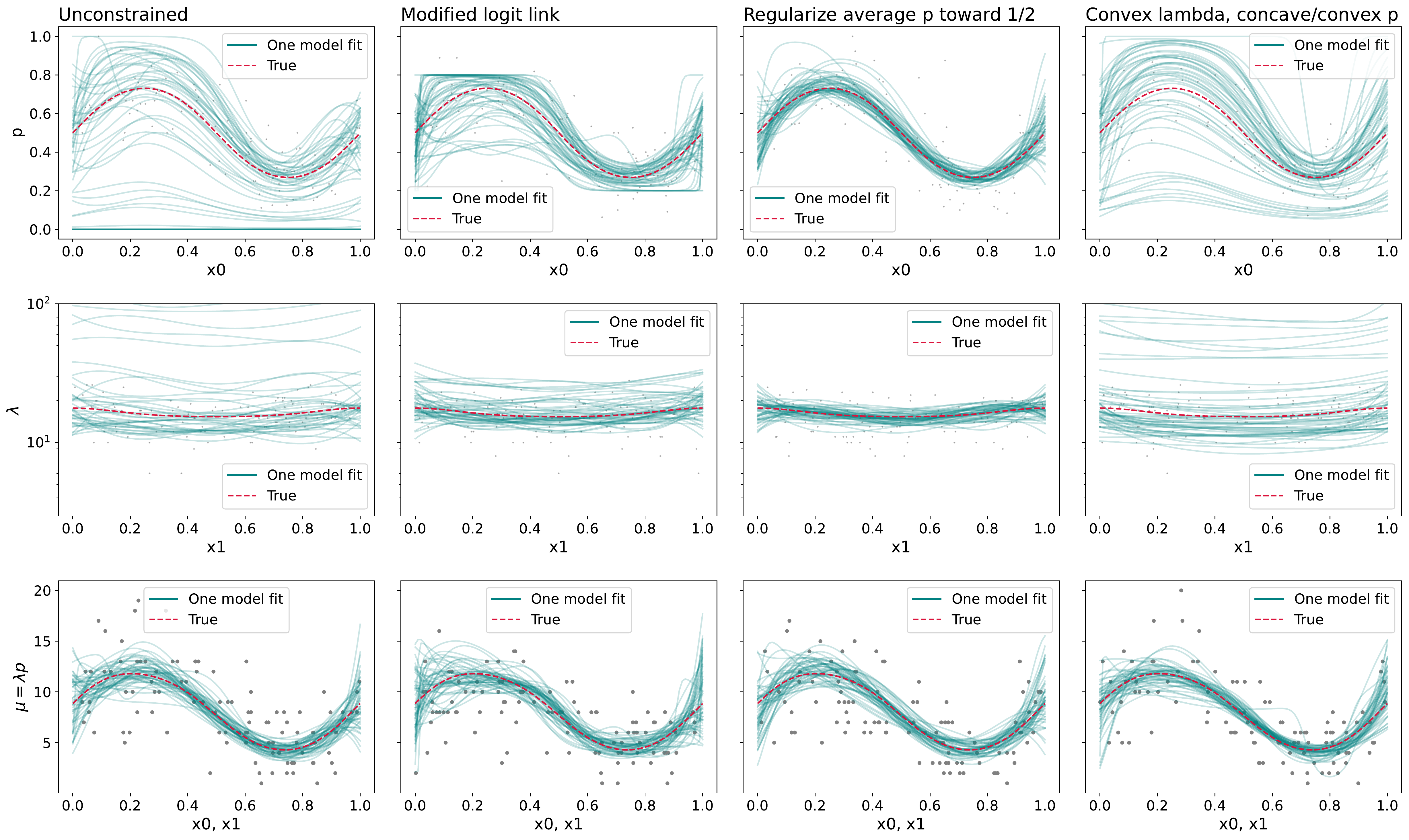}
    \caption{Improvements in fits vs. baseline pogit model (first column) of: modified logit link (second column), prior on $p$ (third column), and convexity constraints (fourth column). All methods easily fit $\mu$ (third row) since it is directly informed by the observations. All three innovations improve fit for $p$ (first row) and $\lambda$ (second row), with the prior on $p$ showing the largest impact in this example.}
    \label{fig:constraintComparisons}
\end{figure}
%\sandy{In figure caption, why do you state "All methods easily fit..." and in the next sentence state that "All three innovations improve..." Is there a difference in meaning between the subjects in these sentences that prevents your using a "they" in the second sentence?}
% ^ Yes, there are FOUR methods in the first sentence, and THREE innovations in the second.

\subsection{Uncertainty quantification and model diagnostics}
\label{sec:UQ}

We use a robust approach to uncertainty quantification that allows estimation of both the uncertainty for $p, \lambda, \mu$ as well as covariate multipliers describing the relationships between these parameters and covariates. Sandwich estimation~\citep{kauermann2001note, wakefield2013bayesian} is robust to model misspecification, which proves to be particularly important for the Pogit model. 
Specifically, our variance-covariance matrix is computed as
\[
V = A^{-1}BA^{-1},
\]
where $A$ is the Hessian of the log likelihood $\mathbb{E}_\theta[\nabla_\theta^2 \ell(x, y | \theta)]$, and $B$ is the Gauss-Newton Hessian approximation $\mathbb{E}_\theta[\nabla_\theta \ell(x, y | \theta) \nabla_\theta \ell(x, y | \theta)^\top]$, both computed at the maximum likelihood estimate by their empirical approximations of the expectations. 
In practice, this approach reports wider uncertainty intervals in the presence of model mis-specification, helping modelers to detect difficult cases. 

\section{Case Studies: Validation on Injury Datasets}
\label{sec:real}
% \textcolor{red}{This intro paragraph needs to be redone to match the new approach.}
% Foreshadow results, what are we trying to show here (homicide is more difficult to model)

We present two case studies exploring the performance of the Pogit model on health-related datasets. Our two case studies, on interpersonal violence and diabetes, illustrate the use of prior knowledge in the form of covariates, constraints and regularization to address the challenges of identifiability and high variance (described in Section~\ref{sec:difficulties}).
In the interpersonal violence study, we estimate the rate of injuries warranting medical
care using data from injuries warranting only \textit{inpatient} medical care,
allowing us to apply the ``under-reporting'' framework. 
For the diabetes study, we estimate the overall rate of medical encounters, again using only inpatient data. 
For each case, we validate our predictions from the Pogit model by comparing them to the total of inpatient \textit{and }outpatient data.  %We use the datasets to illustrate the difficulties of solving the under-reporting problem, and illustrate how in practice using priors and constraints can help resolve these difficulties. 

%\sandy{Please carefully read the preceding paragraph for technical correctness given the many changes I made to it.}

% \subsection{Data Set}
% \textcolor{red}{update to encompass diabetes}

% 

% The under-reporting methodology can be used in settings where only inpatient data is available, and we need to infer total injuries in the absence of reported outpatient data.  
% However, our goal here is to validate new methods, so we consider the data-rich setting where we have both inpatient and outpatient data. This way, when we estimate total injuries using only observed inpatient data only, we can compare our prediction to ground truth using real data. 

%We mimic the underreporting process by imagining we can only observe the inpatient records, but we want to recover the total (inpatient + outpatient) records. 

%In the road injuries data, by contrast, the model is better identified, and the data generating mechanism can be recovered more accurately with less aggressive regularization. \textcolor{red}{Check on this, is it true with how splines work?}

\subsection{Case Study: Interpersonal Violence}
In the International Classification for Diseases 9 (ICD-9) and 10 (ICD-10) codes, injuries are classified by their cause (e.g., interpersonal violence) and/or their nature (e.g.,  traumatic brain injury)\cite{vos2020global}. In addition, they are reported separately based on outpatient or inpatient status. For this case study, we consider all injuries resulting from interpersonal violence and separate them by treatment inside or outside the hospital setting. 

For the validation setting,  $\lambda$ is the true rate of all inpatient and outpatient injuries combined, $p$ is the proportion of injuries that are seen in the inpatient setting, and $\mu$ is the observed rate of injuries in the inpatient category. Our goal is to recover the total rate of injuries due to interpersonal violence \textit{from only inpatient information}.
We use inpatient and outpatient interpersonal violence injury data aggregated at the national level for the US by The Global Burden of Disease study\citep{NAMCS}. 
Only two covariates, age and sex, are available in this dataset. 

Interpersonal violence illustrates an interesting case for our models: the covariates controlling the reporting rate are a strict subset of the covariates controlling the true rate, leading to a loss of identifiability that can be recovered only by using constraints. In particular, as explained below, we model the rate of injury as a function of age and sex  and the probability of inpatient care using only sex. 

% For each cause of injury, we may use covariates to predict the true rate of injury $\lambda$, well as the under-reporting rate $p$. Having the right sets of covariates may be crucial in our ability to disentangle the true rate from the under-reporting $p$. For road injuries and interpersonal violence,

\begin{figure}[h!]
    \centering
    \includegraphics[width=0.5\textwidth]{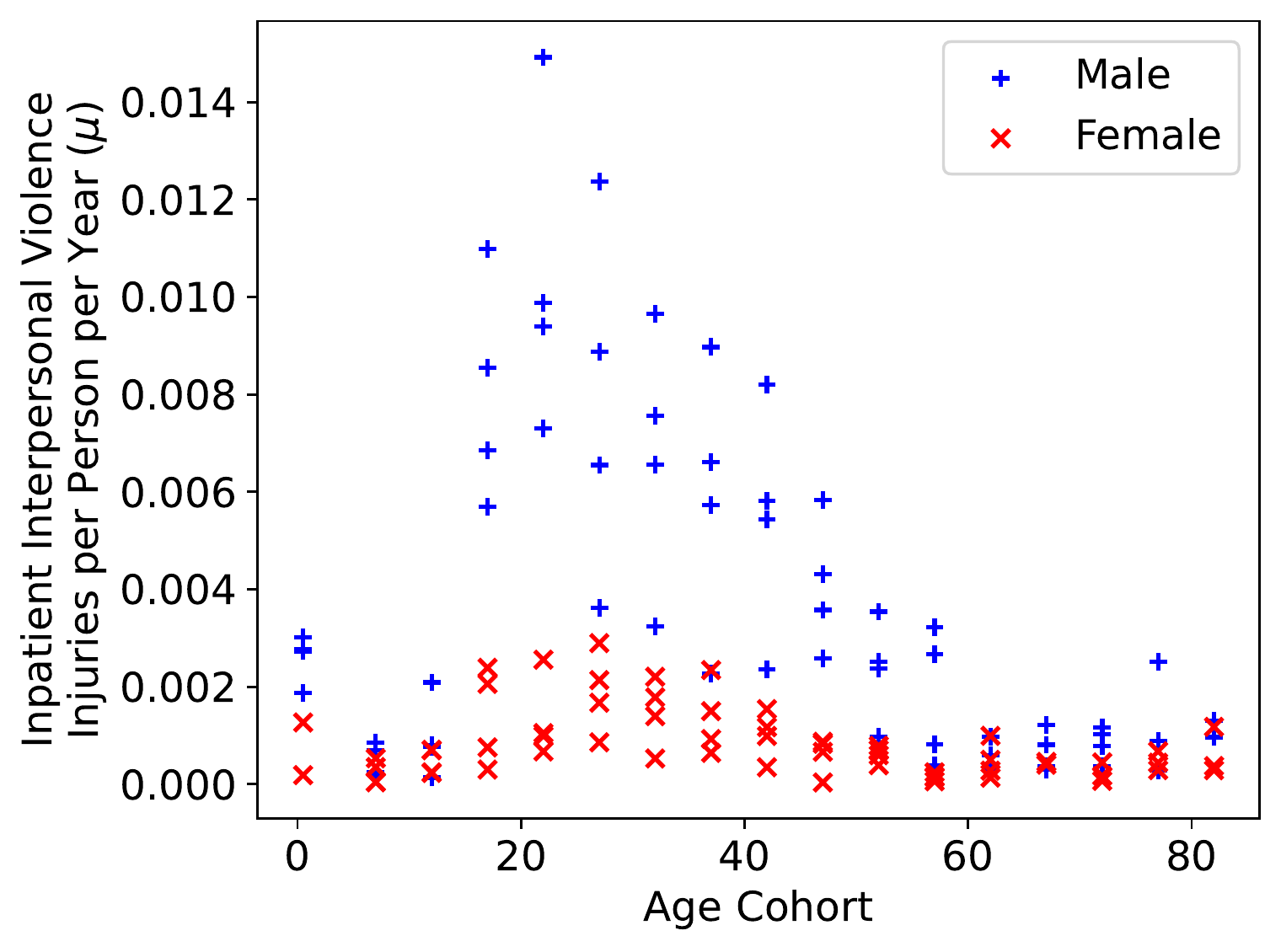}
    \caption{Observed (inpatient) interpersonal violence injuries in the United States, plotted by age/sex, over four five-year periods between 1993 and 2012.}
    \label{fig:obsHomicide}
\end{figure}

Figure \ref{fig:obsHomicide} shows the observed (inpatient) injury rate per person per year, split by age and sex cohorts, for five-year periods between 1993 and 2012. There are clear age and sex effects in the data, with young adult males having the highest rates of injuries requiring inpatient care. 

\begin{figure}[h!]
    \centering
    \includegraphics[width=0.4\textwidth]{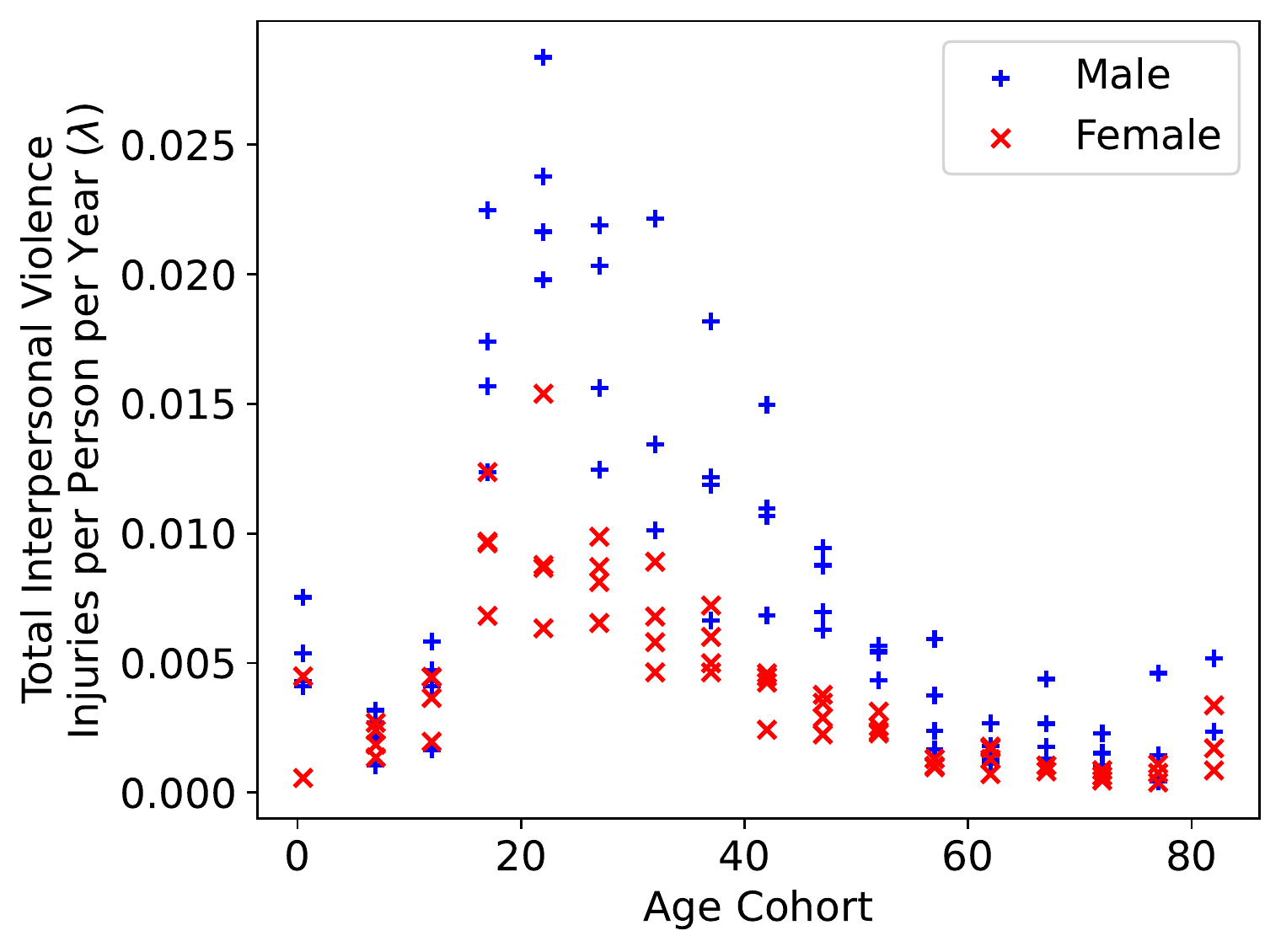}
    \includegraphics[width=0.4\textwidth]{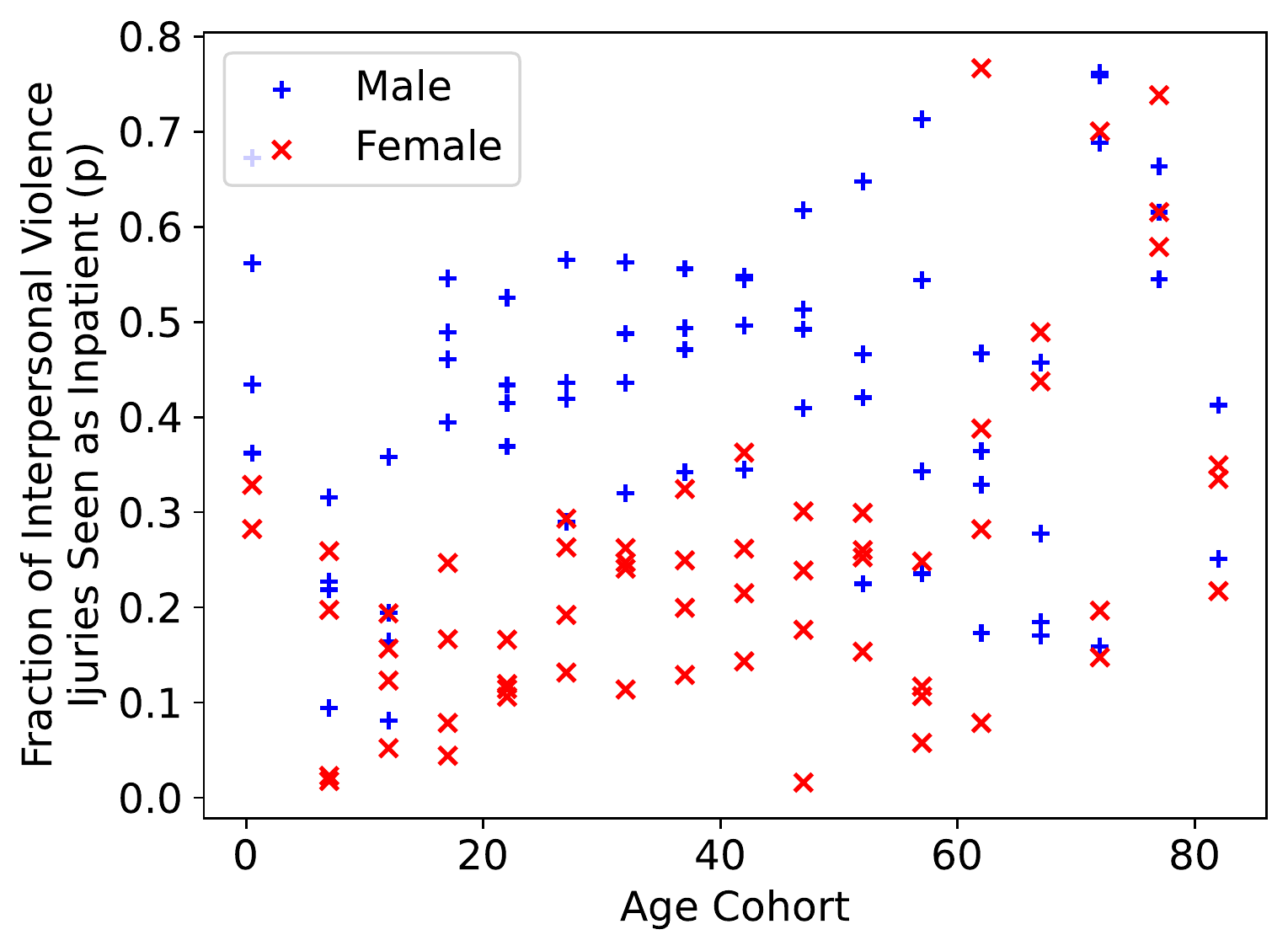}
    \caption{True plots of $\lambda$, the combined rate of inpatient and outpatient injuries (left),  and $p$, the rate at which injuries are treated as inpatient (right).  The data used to make these plots is not provided to the Pogit model, which sees only inpatient data.}
    \label{fig:pLamHomicide}
\end{figure}

Domain knowledge is a critical component of the modeling process, and identifiability of the Pogit model depends on proper modeling choices for $p$ and $\lambda$. In both of our case studies, we use plots of true $p$ and $\lambda$, shown in Figure \ref{fig:pLamHomicide}, to make reasonable choices for their functional forms. Based on these plots, we model $\lambda$ as a spline in age and sex and $p$ as a function of sex alone. This information would not be available to modelers in the real under-reporting setting, who would need domain knowledge to determine the functional forms of $p$ and $\lambda$.

\subsubsection{Modeling $\lambda$ as a function of age and sex.}
We model the true injury rate, $\lambda$, as a function of age $a_i$, sex $s_i$ (coded $0$ for males and $1$ for females) and a fitted intercept: 
\begin{align}
    \lambda_i &= \exp(\beta_{\lambda,0} + \beta_{\lambda,1} s_i + f_\lambda(a_i)).\label{eqn:homLam}
\end{align}
Age enters the model as a cubic spline $f_\lambda$ with a knot at age $15$. The placement of knots can be guided by domain knowledge, e.g., about change points in interpersonal violence based on age. 
% Our model was obtained by looking at the true rate of injuries from combined inpatient and outpatient cohorts, but the same model could be generated from domain knowledge that males have higher injury rates than females, and that the transition between childhood and adulthood is accompanied by a sharp change in rates of interpersonal violence.

\subsubsection{Modeling $p$ as a function of sex alone.}
As Figure \ref{fig:pLamHomicide} shows, the fraction of interpersonal violence requiring inpatient care (the fraction ``reported'') is primarily a function of sex, with a higher rate for males than females. We model this as 
\begin{align}
    p_i &= \frac{\exp(\beta_{p,0} + \beta_{p,1} s_i)}{1 + \exp(\beta_{p,0} + \beta_{p,1} s_i)}. \label{eqn:homP}
\end{align}
% Again, this could be derived by domain knowledge if we believed that the primary factor driving inpatient vs outpatient treatment was the severity of the injury, and if we believed that the severity of injuries varied between the two sexes.

\subsubsection{Constraints and regularization.}
% Because the variables in p are a subset of those in lambda, past work tells us that the model is unidentifiable. 
The covariates modeling $p$ are a subset of those used to model $\lambda$. As discussed by~\cite{papadopoulos2008identification} and in Section \ref{subsec:identifiability}, this overlap in covariates renders the model unidentifiable.
% This means we have to add constraints (males have higher reporting rates than females).
To recover identifiability, we add the constraint that males seek inpatient care at higher rates than females.
%which could be motivated by the assumption that injuries in males are typically more severe than those in females. 
This constrains $\beta_{p,1} > 0$ in Eqn \eqref{eqn:homP} and results in an identifiable model.

% We also find that there is high variance in the fit, so we apply regularization to p.
Even with this constraint, 
%that males receive inpatient care at higher rates than females, 
substantial variance remains in the model predictions for $p$ and $\lambda$. Our approach to further reduce this variance is to add regularization for $p$, pushing $p$ toward $0.5$ using an $\ell_2$-norm penalty on the magnitude of $\beta_{p,0} + \beta_{p,1} s_i$.
%\sandy{Did you mean "pushing"?}
% Lol, yes, thanks

\subsubsection{Results.}\label{sec:homicideResults}
\begin{figure}[h!]
    \centering
    \begin{subfigure}[b]{0.45\textwidth}
         \centering
         \includegraphics[width=\textwidth]{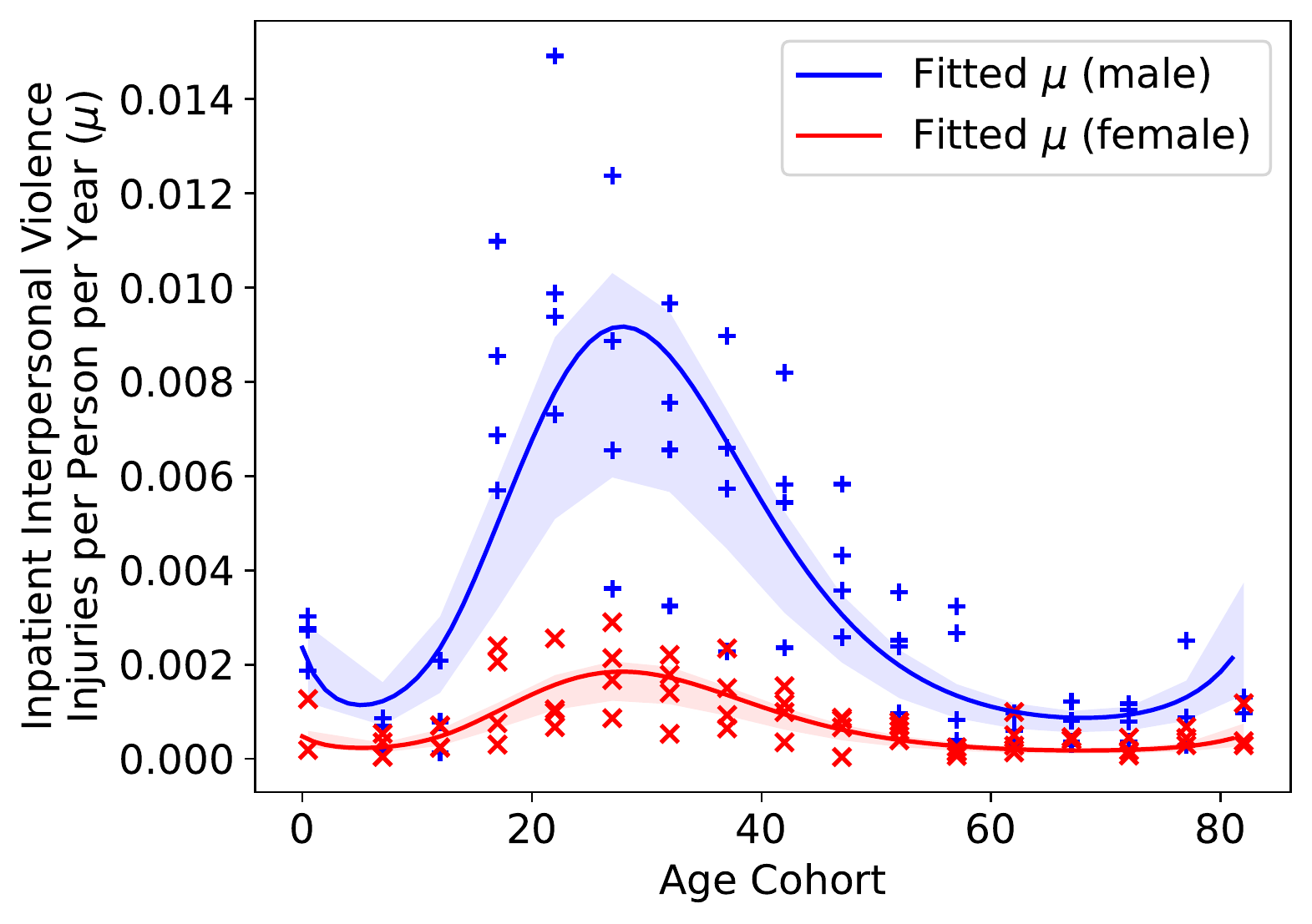}
         \caption{}
         \label{fig:homicideFitted-a}
    \end{subfigure}
    \hfill
    \begin{subfigure}[b]{0.45\textwidth}
         \centering
         \includegraphics[width=\textwidth]{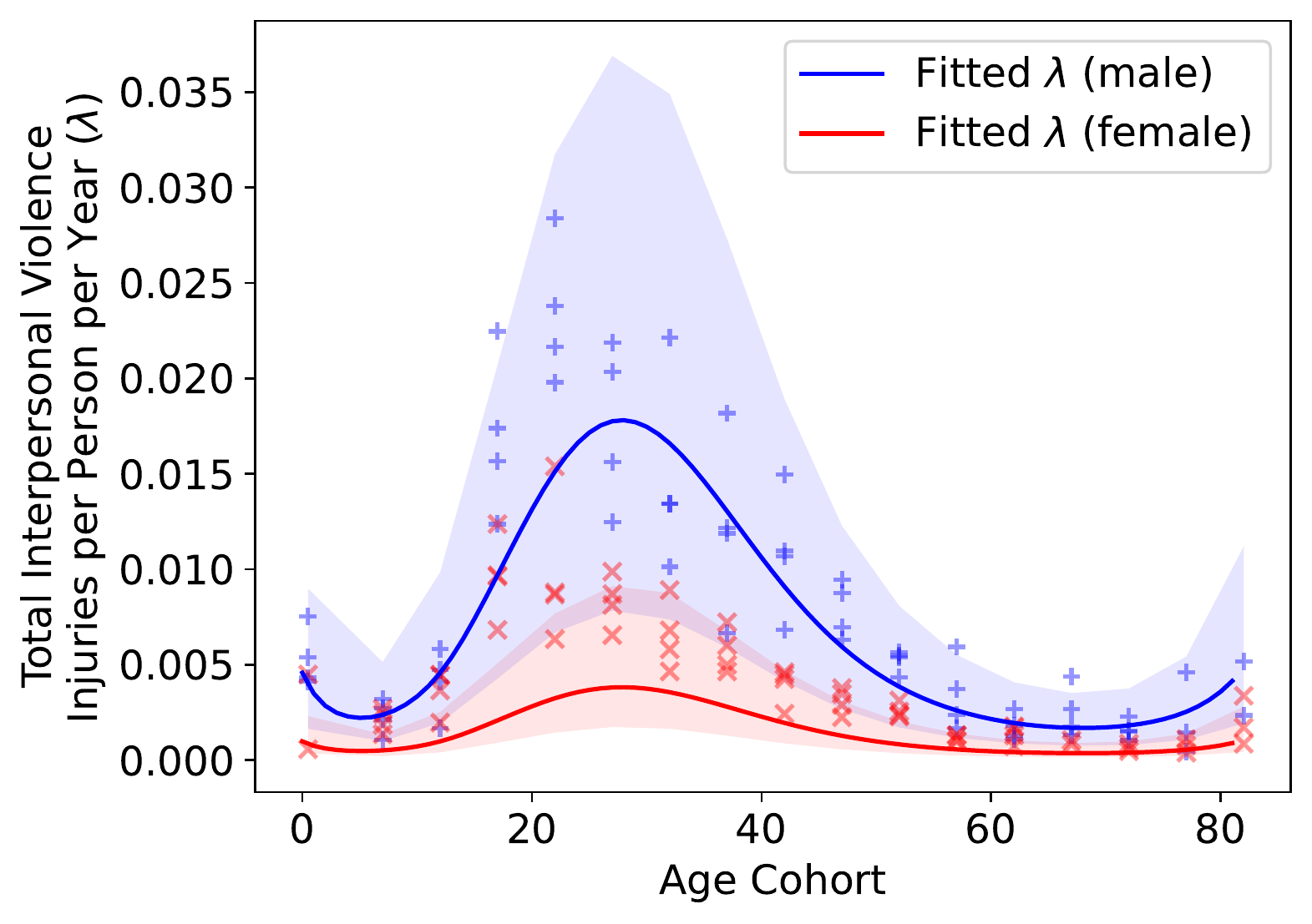}
         \caption{}
         \label{fig:homicideFitted-b}
    \end{subfigure}
    \\
    \begin{subfigure}[b]{0.45\textwidth}
         \centering
         \includegraphics[width=\textwidth]{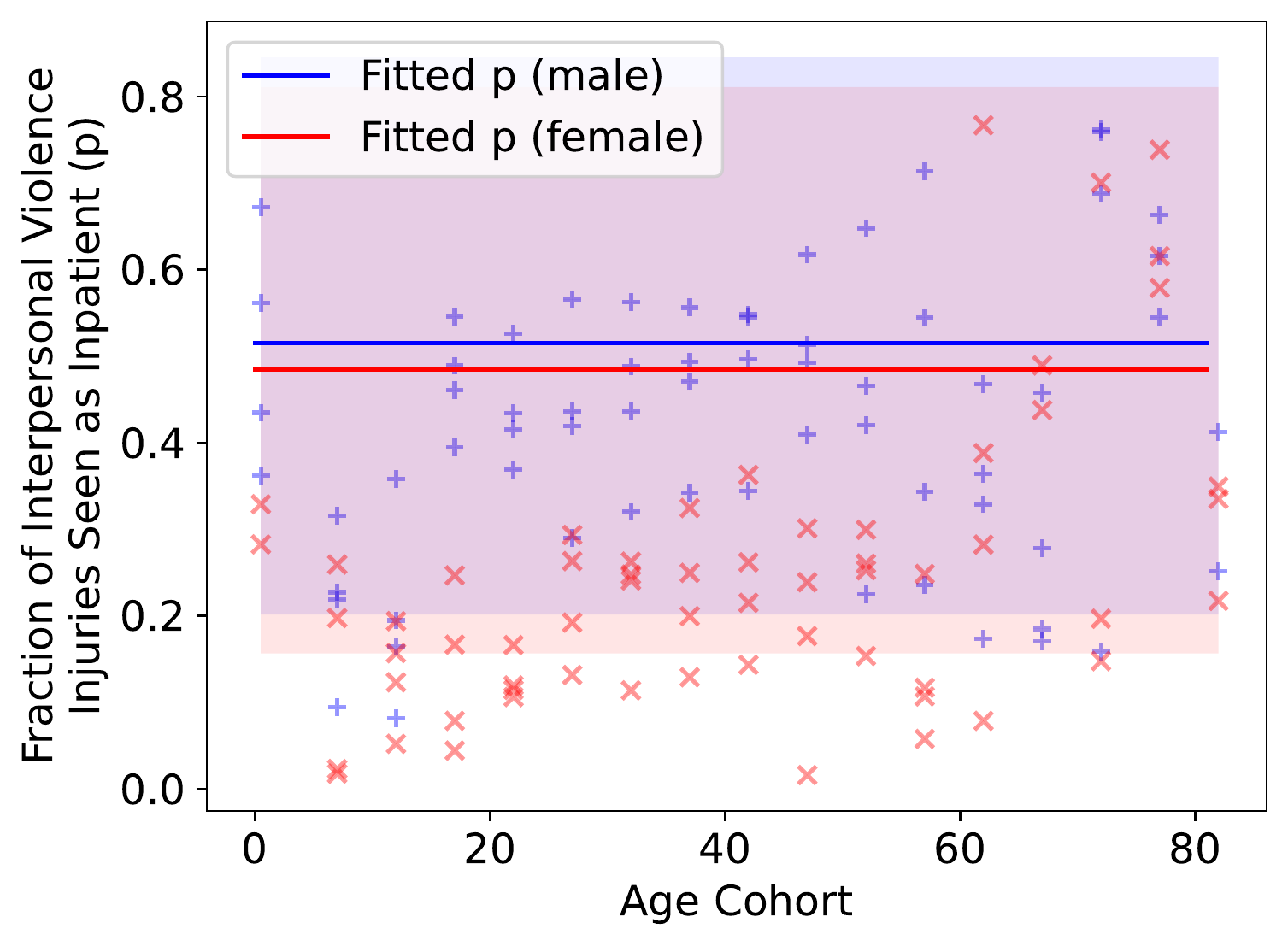}
         \caption{}
         \label{fig:homicideFitted-c}
    \end{subfigure}
    % \hfill
    % \begin{subfigure}[b]{0.45\textwidth}
    %      \centering
    %      \includegraphics[width=\textwidth]{Figs/Homicide/Homicide_blind_pearsonResid.pdf}
    %      \caption{}
    %      \label{fig:homicideFitted-d}
    % \end{subfigure}
    \caption{
    Regularized Pogit fits for the interpersonal violence injuries model using only inpatient data. The estimated total rate $\lambda$ and fraction inpatient $p$ are plotted for each age/sex cohort against validation data not available to the model. Shaded intervals are 90\% confidence intervals computed using sandwich estimation, as described in Section \ref{sec:UQ}.}
    \label{fig:homicideFitted}
\end{figure}

Even with regularization and constraints, the variance of model predictions remains high. Figure \ref{fig:homicideFitted} shows the fitted model and its components $\hat{p}$ and $\hat{\lambda}$. We see that (1) the fitted $\hat{p}$ is higher than the true reporting rate for both sexes, and (2) the variance is so high that the confidence intervals span almost the entire range $[0,1]$, signalling that the problem is difficult. Nonetheless, we can still recover a reasonable estimate for $\hat{\lambda}$, which is typically the more important quantity from a global health perspective. 

We obtain a quantitative comparison of our $\hat\lambda$ estimate vs. baseline estimates using the Akaike Information Criteria (AIC). The AIC is given by $2k - 2\log \hat{\ell}$, where $\ell$ is the likelihood of the model on the fully reported data $Y_i^*$ (see Equation \eqref{eqn:reporting}) and $k$ is the number of parameters. The first model we compare to is the \textit{oracle model} of $\lambda$. The oracle observes the combined inpatient and outpatient data ($Y_i^*$, in the notation of Equation \eqref{eqn:trueEvents}) and fits the Poisson model \eqref{eqn:homLam} to that data. This represents the maximum likelihood fit to $\lambda$ within the model class of Equation \eqref{eqn:homLam}. Our second comparison is to the naive baseline of \textit{ignoring under-reporting}. For this baseline, the Poisson model \eqref{eqn:homLam} is fit to the inpatient data only ($Y_i$ in the notation of Equation \eqref{eqn:reporting}). This baseline, which represents the model that is unaware of the under-reporting problem, will have a low likelihood on the true $Y_i^*$ when the observations are severely under-reported.

Table \ref{table:homicide} shows the AIC values for the three models we consider. Since all three models are of the same parametric form, the $2k$ term acts as a constant offset. We see that the Pogit fit is significantly better compared to ignoring under-reporting  and significantly worse than the oracle fit ($p < 10^{-10}$ with a likelihood ratio test in both cases).

%\footnote{The p-values here are ridiculous, like $10^{-1000}$, it feels kind of ridiculous to even say this.}

%We can compare our estimate of $\hat\lambda$ to various baselines by computing the log likelihood of the reported total number of injuries under each model. Our Pogit model achieves a 3.4x increase in log likelihood compared to the naive model that ignores under-reporting by fitting the Poisson model \eqref{eqn:homLam} to the observed (inpatient) injuries alone. Our model achieves a log likelihood only 2.5x worse than the oracle method that fits the Poisson model \eqref{eqn:homLam} to the \textit{unseen} total injury rate.

%If we use our fitted $\hat\lambda$ to estimate the total number of interpersonal violence injuries, the mean squared error (MSE) for $\log(\hat\lambda)$ is $0.64$, while a baseline predictor that only uses inpatient injuries and fails to account for under-reporting has an MSE of $2.3$ for $\log(\hat\lambda)$.

\begin{table}[h!]
\centering
\caption{AIC of the true interpersonal violence injury rate $\lambda$ over all data points $Y_i^*$, reported under three different models: the ``oracle'' Poisson fit of the injury rate to model \eqref{eqn:homLam} if we could observe the number of total injuries $Y_i^*$; the Poisson portion \eqref{eqn:homLam} of the Pogit model fit to inpatient injuries $Y_i$; and the naive baseline that ignores under-reporting by fitting the Poisson model \eqref{eqn:homLam} to the observed inpatient data $Y_i$.}
\begin{tabular}{|c r r r|} 
 \hline
  & Oracle Fit & Pogit Fit & Ignoring Under-reporting \\ [0.5ex] 
 \hline\hline
 AIC & 18000 & 62000 & 156000 \\ [1ex] 
 \hline
\end{tabular}
\label{table:homicide}
\end{table}

\subsection{Case Study: Diabetes Care}
\begin{figure}[h!]
    \centering
    \begin{subfigure}[b]{0.45\textwidth}
         \centering
         \includegraphics[width=\textwidth]{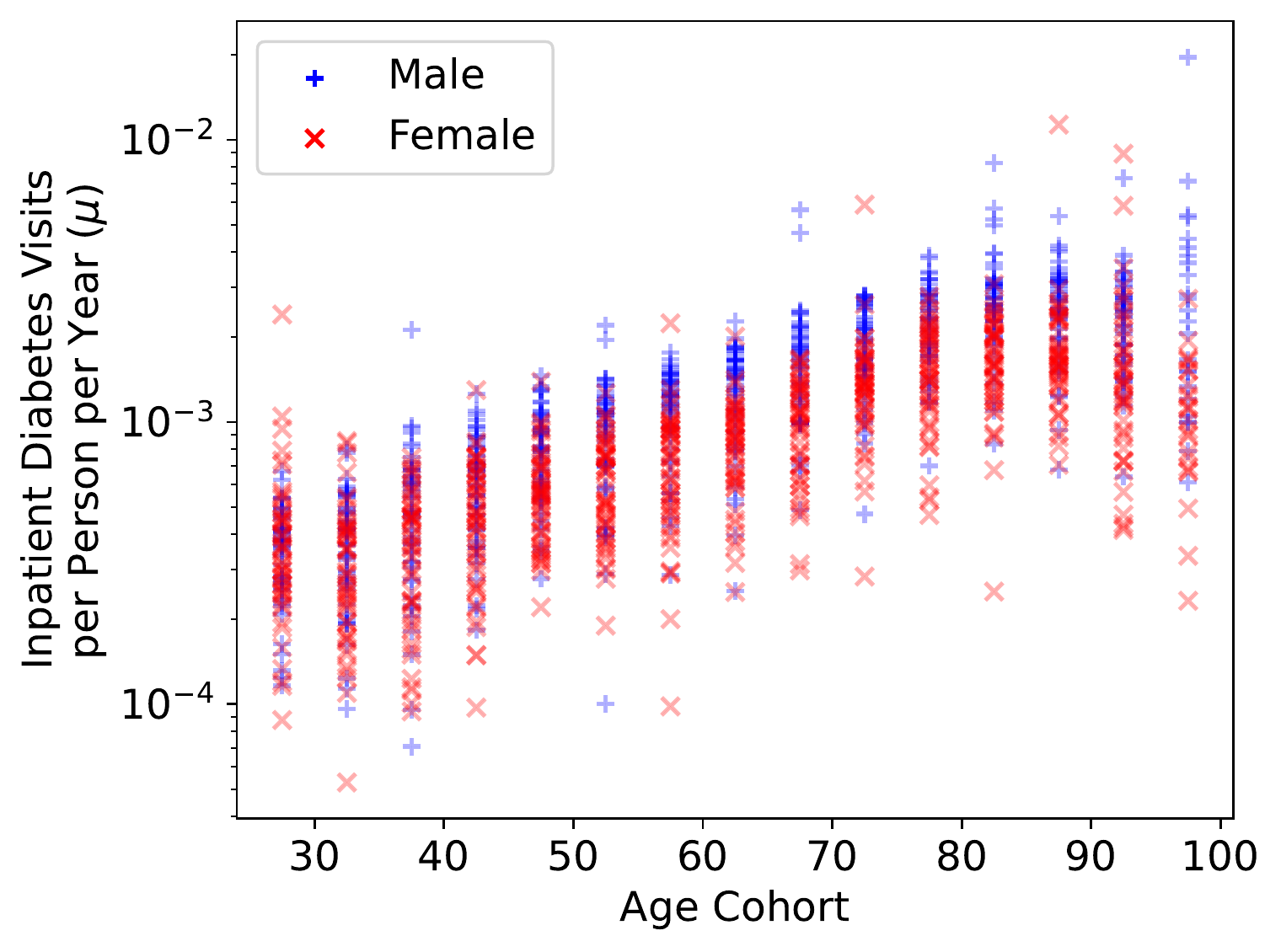}
         \caption{}
         \label{fig:diabetes-muObs-age}
    \end{subfigure}
    \hfill
    \begin{subfigure}[b]{0.45\textwidth}
         \centering
         \includegraphics[width=\textwidth]{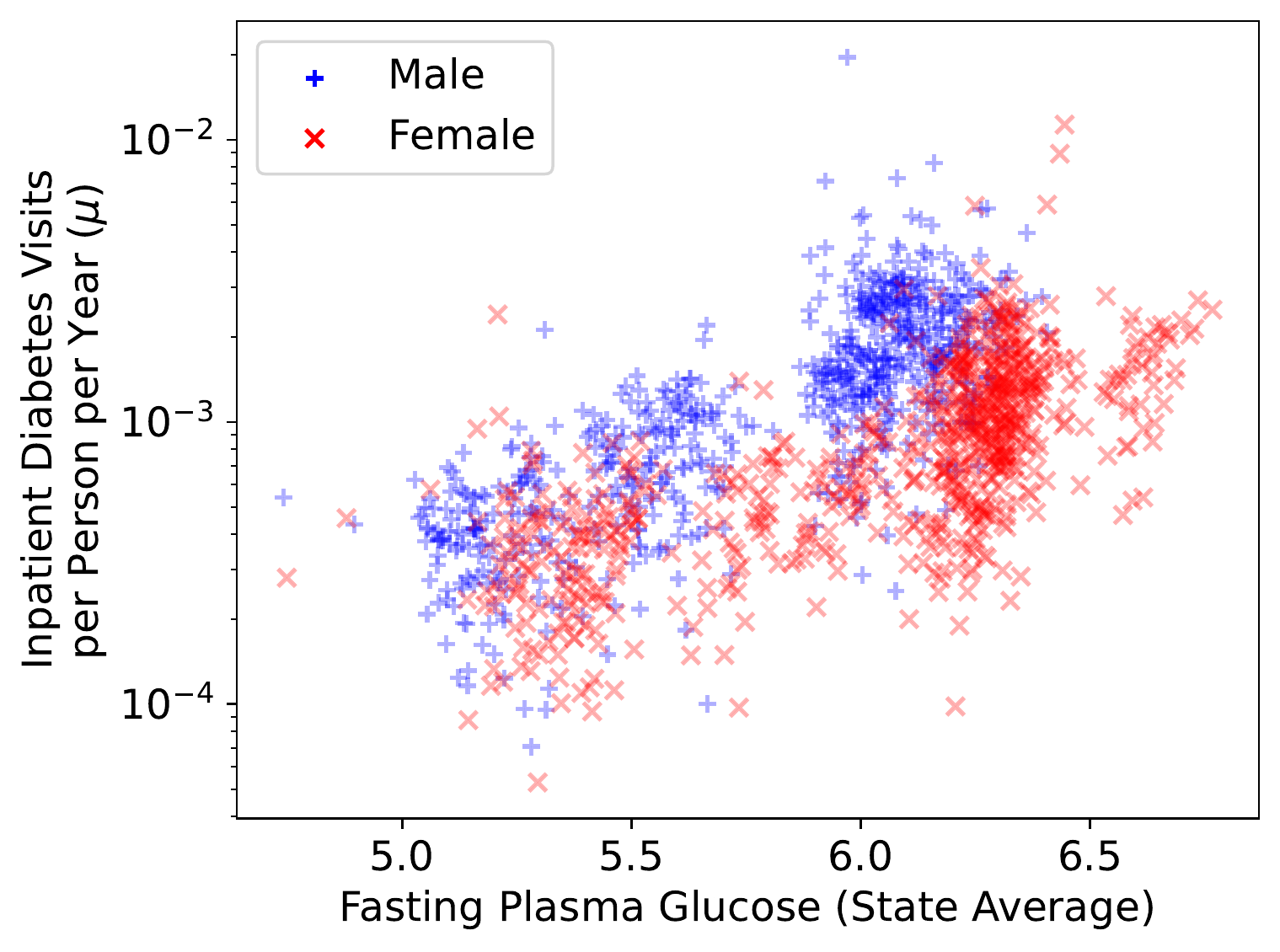}
         \caption{}
         \label{fig:diabetes-muObs-fpg}
    \end{subfigure}
    \caption{Inpatient diabetes care for age/sex (left) and cohort-average FPG (right) in 2019.}
    \label{fig:diabetesMu}
\end{figure}

\begin{figure}[h!]
    \centering
    \begin{subfigure}[b]{0.45\textwidth}
         \centering
         \includegraphics[width=\textwidth]{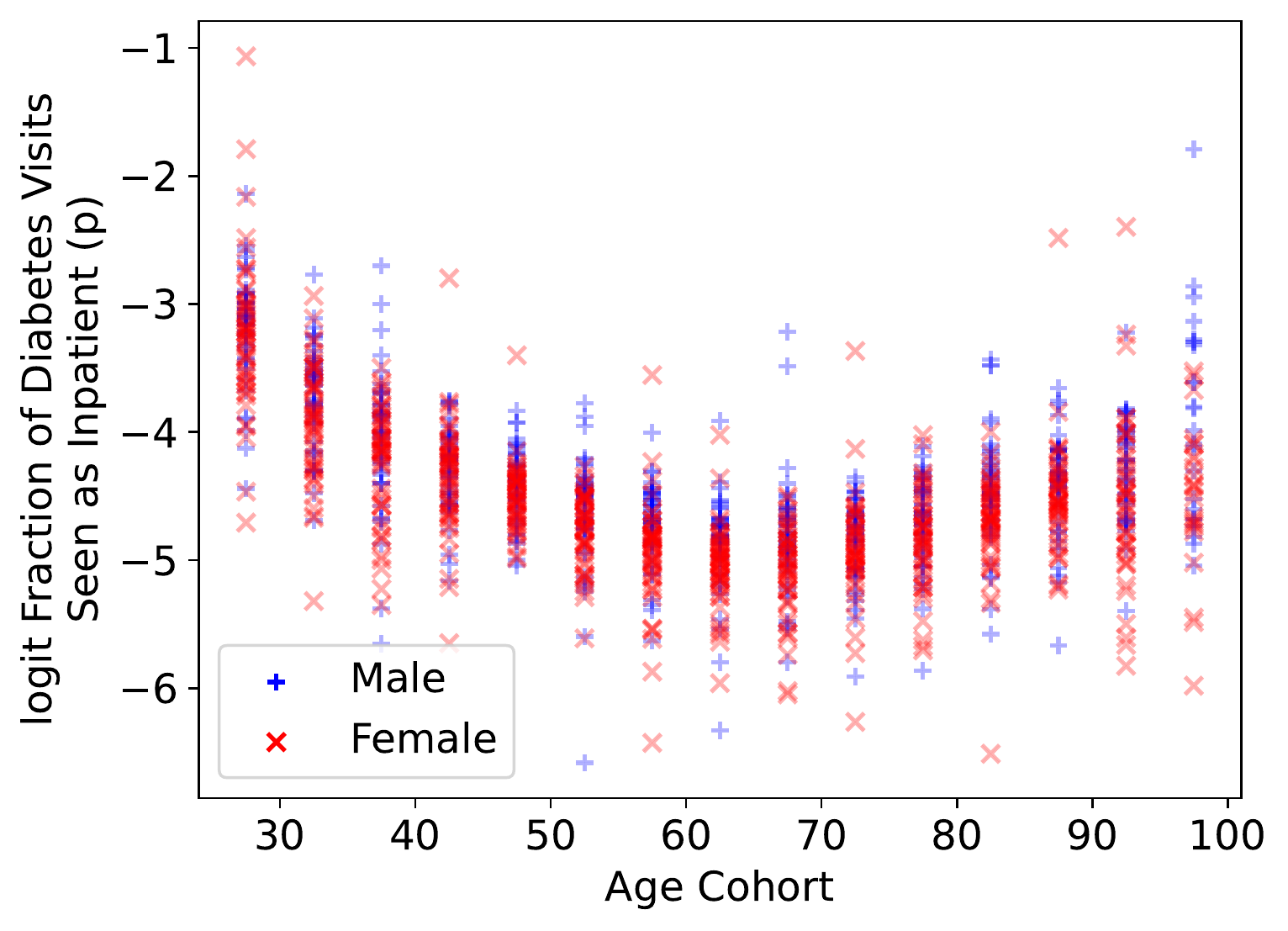}
         \caption{}
         %\label{fig:diabetes-pObs}
    \end{subfigure}
    \hfill
    \begin{subfigure}[b]{0.45\textwidth}
         \centering
         \includegraphics[width=\textwidth]{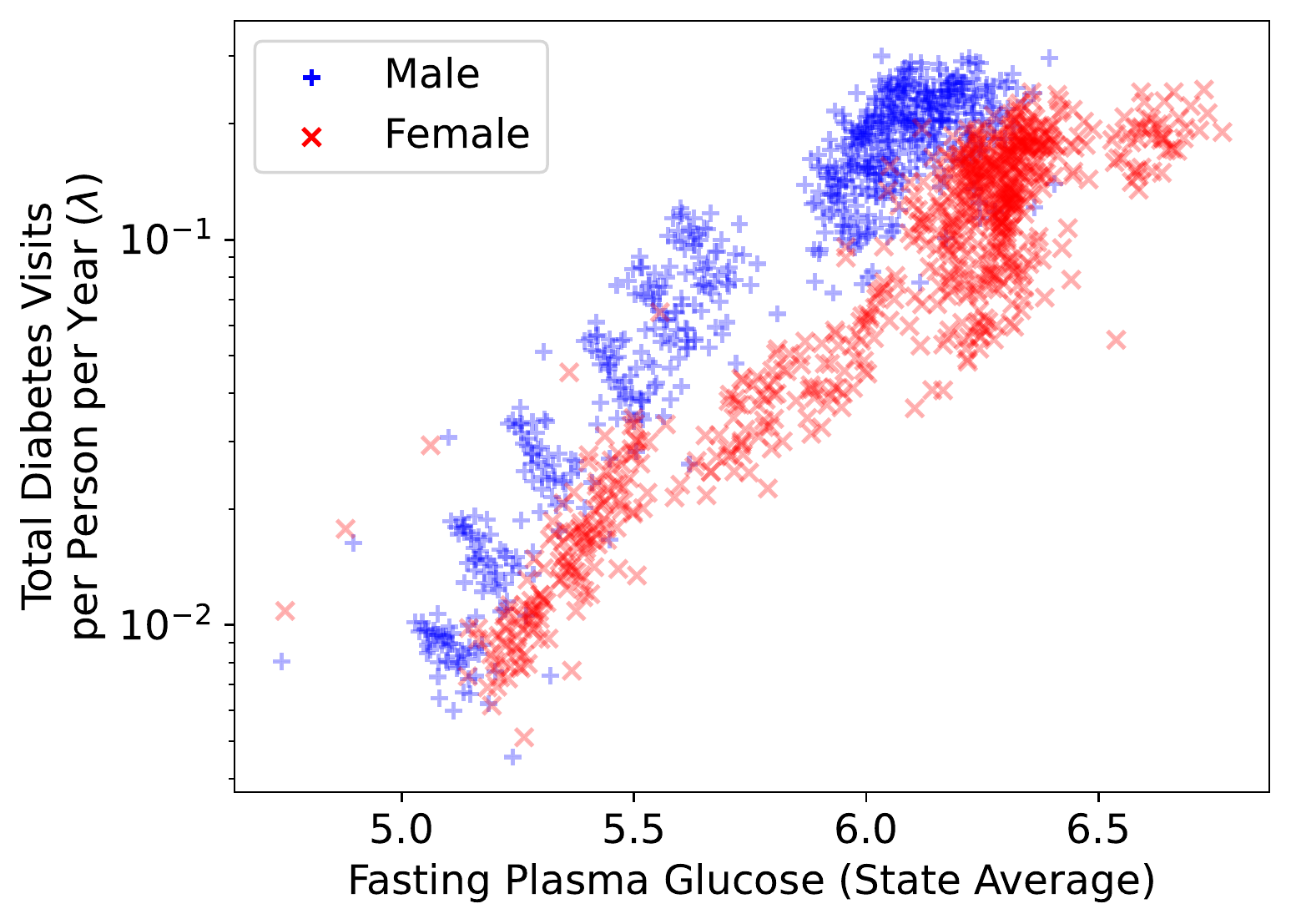}
         \caption{}
         %\label{fig:diabetes-lamObs}
    \end{subfigure}
    \caption{True fraction of inpatient visits as a function of age and sex for 2019 (left). Total diabetes care as a function of population average FPG and sex for 2019 (right).}
    \label{fig:diabetesPLamObs}
\end{figure}

In the second case study, we apply the Pogit model to estimate the rate of diabetes care (inpatient \textit{and} outpatient visits) having observed \textit{only inpatient} visits.  We use MarketScan healthcare claims data that is processed for use in the Global Burden of Disease Study \citep{marketscan, vos2020global, NHDS}. The data is aggregated at the age and sex level for each US state for the year 2019; the state-level aggregation lets us use a richer set of covariates to model $p$ and $\lambda$. Figure \ref{fig:diabetesMu} shows the rate of inpatient diabetes cases per person per year across all fifty states, as a function of age, sex, and population average fasting plasma glucose (FPG) for each state/age/sex cohort. Both age and population average FPG correlate positively with diabetes inpatient admissions. We parametrize models for $p$ and $\lambda$ based on the $p$ and $\lambda$ plots shown in Figure \ref{fig:diabetesPLamObs}.
%which would typically be unavailable to the modeler. 

\subsubsection{Modeling $\lambda$ as a function of FPG and sex.}
We model the true rate of diabetes care, $\lambda$, as a function of sex $s_i$ and the state average FPG $g_i$ with a fitted intercept:
\begin{align}
    \lambda_i &= \exp(\beta_{\lambda, 0} + \beta_{\lambda, 1} s_i + f_\lambda(g_i)).\label{eqn:diabetes_lam}
\end{align}
FPG enters the model as a quadratic spline $f_\lambda$.

\subsubsection{Modeling $p$ as a function of age.}
The observed rate of inpatient diabetes care is driven by the true rate of diabetes care and by the fraction of care that is treated in the inpatient vs. outpatient setting. Based on Figure \ref{fig:diabetesPLamObs}, we 
model $p$ as a quadratic spline in age:
\begin{align}
    p_i &= \frac{\exp(\beta_{p,0} + f_p(a_i))}{1+\exp(\beta_{p,0} + f_p(a_i))}.\label{eqn:diabetes_p}
\end{align}
We apply several constraints and regularizers on $p$ to reduce the variance of our estimate. First, we enforce that $p$ decreases from age 25 to age 60. Second, we apply a quadratic regularization of the average fitted $p$ toward its true average, $0.01$. Including this side information improves the model fits for both $p$ and $\lambda$.

\subsubsection{Results}
\begin{figure}[h!]
    \centering
    \begin{subfigure}[b]{0.45\textwidth}
         \centering
         \includegraphics[width=\textwidth]{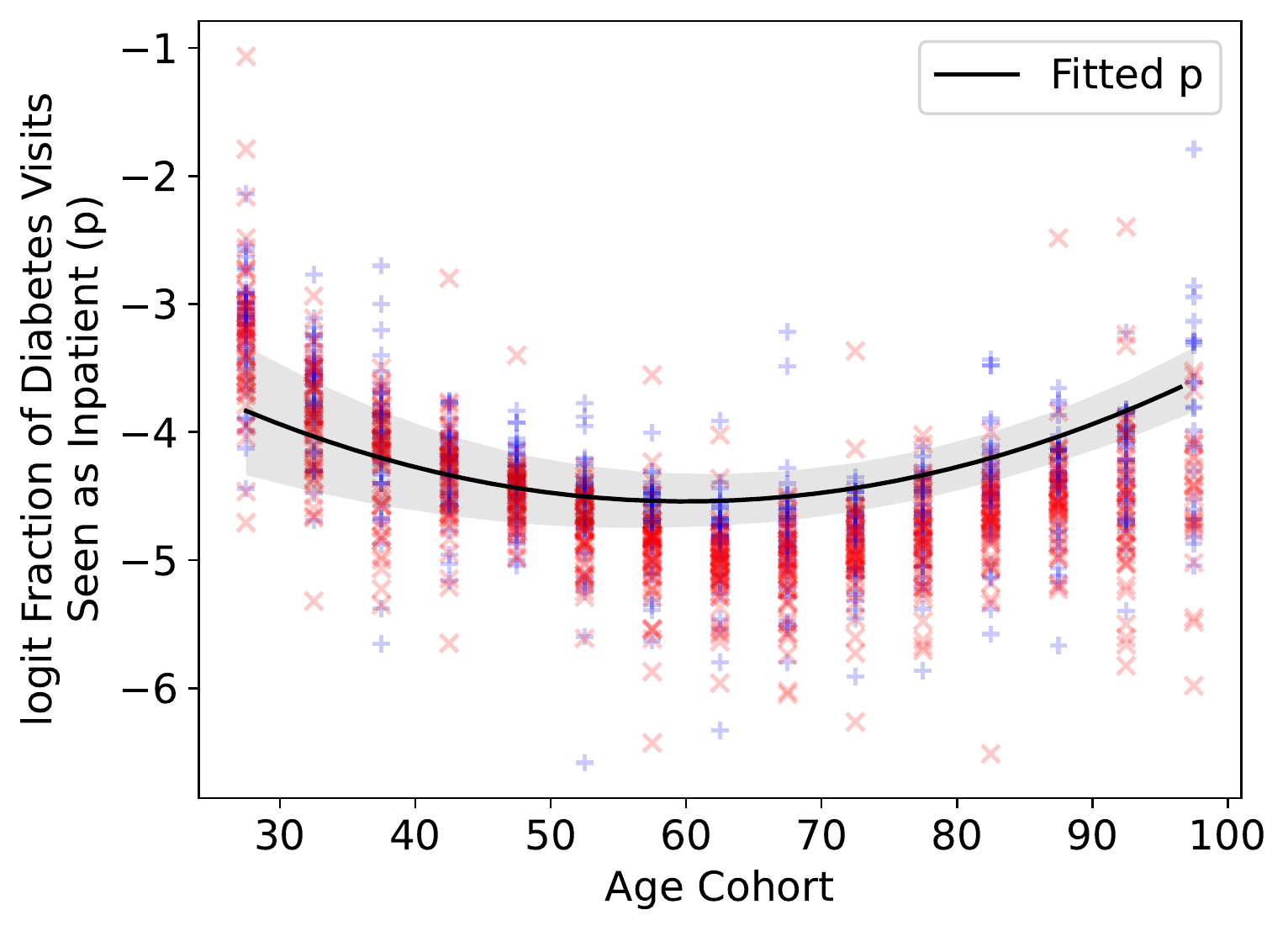}\\
         \includegraphics[width=\textwidth]{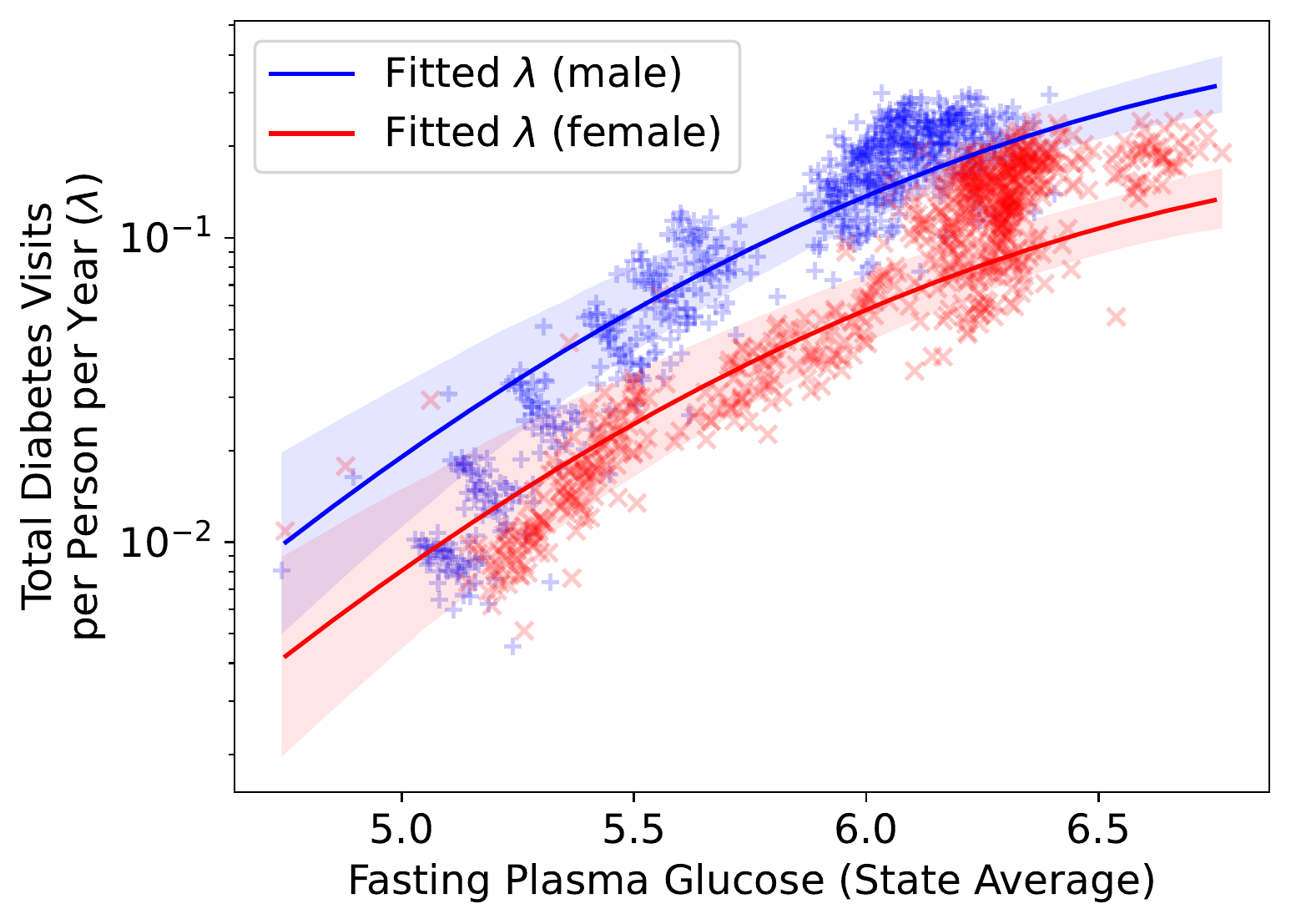}
         \caption{}
         %\label{fig:diabetes-pObs}
    \end{subfigure}
    \hfill
    \begin{subfigure}[b]{0.45\textwidth}
         \centering
         \includegraphics[width=\textwidth]{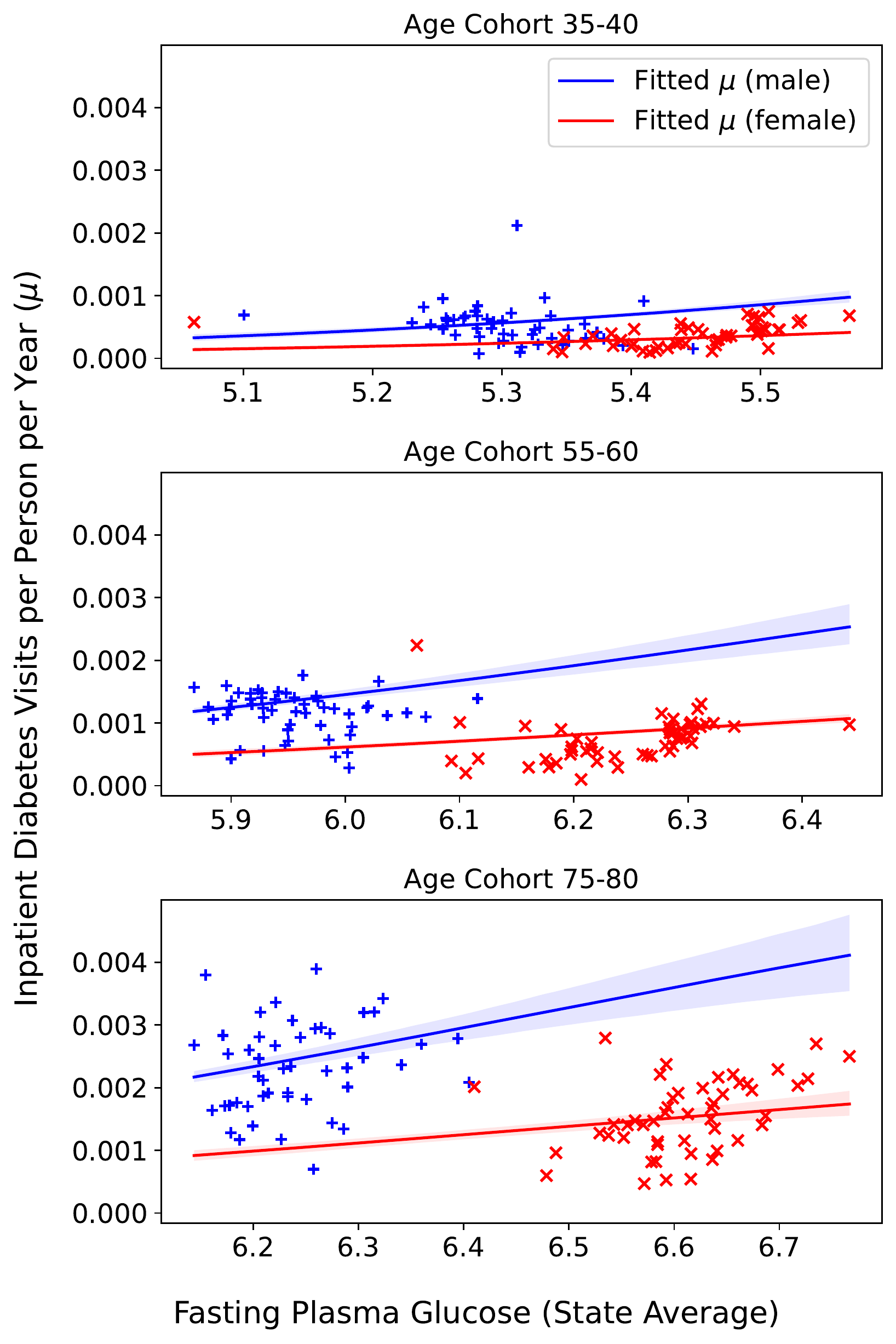}
         \caption{}
    \end{subfigure}
    % \\
    % \begin{subfigure}[b]{0.45\textwidth}
    %      \centering
    %      \includegraphics[width=\textwidth]{Figs/Diabetes/Diabetes-pearsonResid-age.pdf}
    %      \caption{}
    %      \label{fig:diabetes-residuals-age}
    % \end{subfigure}
    % \hfill
    % \begin{subfigure}[b]{0.45\textwidth}
    %      \centering
    %      \includegraphics[width=\textwidth]{Figs/Diabetes/Diabetes-pearsonResid-fpg.pdf}
    %      \caption{}
    %      \label{fig:diabetes-residuals-fpg}
    % \end{subfigure}
    \caption{Regularized Pogit fits for the diabetes care model using only inpatient data. The estimated total rate of care $\lambda$ and fraction of inpatient care $p$ are plotted for each age/sex cohort against validation data that was not available to the model. The fitted inpatient care rate $\mu$ is a function of age, sex and state average FPG, so we provide three plots for different age cohorts. Shaded intervals are 90\% confidence intervals computed using sandwich estimation, as described in Section \ref{sec:UQ}.}
    \label{fig:diabetesFitted}
\end{figure}

We fit the model described in \eqref{eqn:diabetes_lam} and \eqref{eqn:diabetes_p} to the diabetes inpatient care data. Figure \ref{fig:diabetesFitted} shows the results. Both fitted $\hat{p}$ and $\hat\lambda$ capture the important properties of their respective processes: $p$ is convex in age, while $\lambda$ is concave in FPG and shows a higher rate for males than females. 

Following the methodology of Section \ref{sec:homicideResults}, we quantitatively evaluated the fitted model of the combined rate of care, $\lambda$, using the AIC. Table \ref{table:diabetes} shows the AIC of the Pogit model, the oracle with access to both inpatient and outpatient data, and the naive baseline that ignores under-reporting. 
The Pogit fit is significantly better than the naive baseline, but not as good as the model that observes both inpatient and outpatient data ($p < 10^{-10}$ with a likelihood ratio test in both cases). The Pogit model outperforms the naive baseline by a wider margin on the diabetes data than on the interpersonal violence study because the reporting rate is lower for diabetes, so the penalty for ignoring under-reporting is higher.

\begin{table}[h!]
\centering
\caption{AIC of the true diabetes visit rate $\lambda$ over all data points $Y_i^*$, reported under three different models: the ``oracle'' Poisson fit of the visit rate to model \eqref{eqn:diabetes_lam} if we could observe the number of total visits $Y_i^*$; the Poisson portion of the Pogit model fit on only inpatient visits $Y_i$; and the naive baseline that ignores under-reporting by fitting the Poisson model \eqref{eqn:diabetes_lam} to the observed inpatient data $Y_i$.}
\begin{tabular}{|c r r r|} 
 \hline
  & Oracle Fit & Pogit Fit & Ignoring Under-reporting \\ [0.5ex] 
 \hline\hline
 AIC & 450 & 749 & 37400 \\ [1ex] 
 \hline
\end{tabular}
\label{table:diabetes}
\end{table}

%LRT test p-values are $10^{-65}$ for Pogit vs oracle, and $10^{-8000}$ for Pogit vs ignoring under-reporting. Note that the naive baseline is even more separated from the Pogit than in the homicide example because the reporting rate $p$ is much lower for diabetes.
\section{Discussion}
\label{sec:discussion}

In this paper, we presented theoretical challenges in modeling under-reported count data using the Pogit model. We showed how priors and constraints can help resolve these issues and used real-world data from the Global Burden of Disease study to validate our approach. We found that the proposed formulation enables successful estimation, provided sufficient prior information can be specified. Examples of such information include aggregate measures (such as national reporting rate), shape of the relationships, and prior values for specific datapoints and covariate values.  The tools used to create the results and test the methods are available in a publicly accessible repository. 

The approach and analysis in this paper focused on the Pogit model. Future analysis and extensions can be made by considering other count models that better account for over-dispersion. A potential challenge for extensions is that additional flexibility may exacerbate the difficulty of the deconvolution problem.

Another interesting direction for future work is to use the approach developed here to aid in decision making about the kinds of data or information in which to invest. Specifically, decisions to obtain new data sources or conduct additional studies can use the available package to evaluate the type of information most effective for minimizing robust uncertainty estimates. A rigorous framing of this idea is left to future work. 

\section{Software}
\label{sec:software}
The techniques described in this paper have been implemented in the Python package \texttt{Regmod}, available on GitHub at \url{https://github.com/ihmeuw-msca/regmod}. Tutorials are available on GitHub at \url{https://github.com/ihmeuw-msca/underreporting}.

% \section{Supplementary Material}
% \label{sec:supp}

% %Supplementary material is available online at
% %\url{http://biostatistics.oxfordjournals.org}.

% % Sample from https://academic.oup.com/biostatistics/pages/General_Instructions
% % under "stylistic requirements"
% The reader is referred to the online Supplementary Materials for technical appendices.

\section*{Acknowledgments}
The authors would like to thank Emily Johnson for providing the diabetes data set, Madeline Moberg and Erin Hamilton for providing and explaining the road injuries data set, and Dr. Liane Ong for providing expert insight into covariates to use for the diabetes model. We are also grateful to Sandy Kaplan for her review of the manuscript.

{\it Conflict of Interest}: None declared.

\bibliographystyle{biorefs}
\bibliography{refs}

% Figures and tables go at the end of the document, I guess?
\newpage
\appendix 

\section{Proof of the Estimation Error Lower Bound}
\label{app:lowerBound}

In this section, we present proofs of lemmas and theorems in the paper. 
%We begin by formally stating the setting of the analysis. 

\begin{setting}[Two-covariate Pogit Model]\label{set:twoCovPogit}
For $i=1,2,\ldots, n$, let covariates $x_{p, i}$ and $x_{\lambda, i}$ be drawn independently according to
\begin{align}
    x_{\lambda, i} &\sim \mathcal{N}(\mu_\lambda, \sigma_\lambda^2)\\
    %x_{p, i} &\sim \mathcal{D}_p \qquad \text{s.t.}~\mathbb{P}_{\mathcal{D}_p}(x) =\mathbb{P}_{\mathcal{D}_p}(-x) 
    x_{p, i} &\sim \mathcal{N}(0, \sigma_p^2)
\end{align}
and let $Y_i$ be drawn according to
\begin{align}
    Y_i &\sim \text{Poi}\left( e^{x_{\lambda, i}\theta_\lambda} \frac{\exp\left( x_{p, i}\theta_p \right)}{1 + \exp\left( x_{p, i}\theta_p \right)} \right),
\end{align}
where $\theta_\lambda, \theta_p \in [C_l, C_u]$ for constants $C_l, C_u \in \mathbb{R}$. The existence of lower and upper bounds on the parameters is needed to prove certain regularity conditions about the maximum likelihood estimator, but the bounds can be chosen such that they are never attained in practical settings.
\end{setting}

\section{Proof of Lemma~\ref{lemma:mleNormConditions}}
To prove this claim, it suffices to show that the following regularity conditions are satisfied:
%\footnote{From http://www.biostat.jhsph.edu/bstcourse/bio771/sec8b.pdf, although not sure if there's something else we can cite}
\begin{enumerate}
    \item $\theta_0$ is \textit{identified}, in the sense that if $\theta\neq \theta_0$ and $\theta\in\Theta$, then $\ell(x,y|\theta) \neq \ell(x,y|\theta_0)$ with respect to the dominating measure $\mu$.
    \item $\mathbf{\theta_0}$ lies in the interior of $\Theta$, which is assumed to be a compact subset of $\mathbb{R}^2$.
    \item $\log \ell(x, y|\theta)$ is continuously differentiable at each $\theta\in\Theta$ for all $x, y \in \mathcal{X}\times \mathcal{Y}$ (a.e. will suffice).
    \item {$|\log \ell(x, y | \theta)| \leq d(x, y)$ for all $\theta\in\Theta$ and $\E_{\theta_0}[d(X,Y)] < \infty$.}
    \item $\ell(x, y | \theta)$ is twice continuously differentiable, and $\ell(x, y | \theta) > 0$ in a neighborhood, $\mathcal{N}$, of $\theta_0$.
    \item $||\tfrac{\partial \ell(x, y | \theta)}{\partial \theta}|| \leq e(x, y)$ for all $\theta\in \mathcal{N}$ and $\int e(x, y) d\nu(x, y) < \infty$.
    \item Defining the score vector
    \[
    \psi(x,y | \theta) = (\partial \log \ell(x,y|\theta) / \partial \theta_1, \ldots, \partial \log \ell(x,y | \theta) / \partial \theta_k)'
    \]
    then %we assume that 
    $I(\theta_0) = \E_{\theta_0} [\psi(X,Y|\theta_0)\psi(X,Y|\theta_0)']$ exists and is non-singular.
    \item {$|| \tfrac{\partial^2 \log \ell(x,y|\theta)}{\partial \theta \partial \theta'}|| \leq f(x,y)$ for all $\theta\in \mathcal{N}$ and $\E_{\theta_0} [f(X,Y)] < \infty$.}
    \item $|| \tfrac{\partial^2 \ell(x,y|\theta)}{\partial \theta \partial \theta'}|| \leq g(x,y)$ for all $\theta\in \mathcal{N}$ and $\int g(x,y) d\nu(x,y) < \infty$.
\end{enumerate}

We first introduce the likelihood decomposition, which we will use repeatedly in the following proofs:
\begin{equation}
    \label{eq:likelihood_demp}
    \ell(x, y | \theta) = \ell(y | x, \theta) \ell(x) = \ell(y|\mu(x, \theta)) \ell(x)
\end{equation}
And for log likelihood:
\begin{equation}
    \label{eq:log_likelihood_demp}
    \log \ell(x, y|\theta) = \log \ell(y|\mu(x, \theta)) + \log \ell(x).
\end{equation}

{\bf Condition (1)}. 
Identifiability holds because the covariates for $p$, $X_p$, are independent from the covariates for $\lambda$, $X_\lambda$ (as shown in \cite{papadopoulos2008identification}).

{\bf Condition (2)} holds by the assumption given in Setting \ref{set:twoCovPogit}.

{\bf Condition (3)}. From \eqref{eq:log_likelihood_demp} and the fact that $\log \ell(y| \mu)$ is continuously differentiable in $\mu$ and $\mu$ is continuously differentiable in $\theta$, we know that  $\log(x, y|\theta)$ is continuously differentiable in $\theta$.

{\bf Condition (4).}
Since $\Theta$ is a compact set, for every $(x, y)$ we can attain the maximum and minimum value of $\log \ell(x, y|\theta)$ with respect to $\theta$:
\[
\theta_{\max}(x, y) = \max_{\theta \in \Theta} \log \ell(x, y|\theta), \quad
\theta_{\min}(x, y) = \min_{\theta \in \Theta} \log \ell(x, y|\theta)
\]
And we could write out our upper bound function $d$ in terms of $\theta(x, y)$:
\[
d(x, y) = \max\{\left.\log \ell(x, y|\theta)\right|_{\theta=\theta_{\max}(x,y)},
-\left.\log \ell(x, y|\theta)\right|_{\theta=\theta_{\min}(x,y)}
\}
\]
For simplicity, we use $\theta(x,y)$ to ambiguously denote either $\max$ or $\min$ branch.

The expectation in the condition (6) can be written as
\begin{align*}
    &\int_x \sum_{y=0}^\infty \left.\log \ell(x, y|\theta)\right|_{\theta=\theta(x,y)}f_Y(y) f_X(x)dx\\
    =&\int_x \sum_{y=0}^\infty \left.\left[\log \ell(y|\mu(x, \theta)) + \log \ell(x)\right]\right|_{\theta=\theta(x,y)}f_Y(y) f_X(x)dx\\
    =&\int_x \log \ell(x) f_X(x)dx + \int_x \sum_{y=0}^\infty \left.\log \ell(y|\mu(x, \theta))\right|_{\theta=\theta(x,y)}f_Y(y) f_X(x)dx
\end{align*}
Here, we argue that since Poisson model satisfies the regularity condition, we know that 
\[
\sum_{y=0}^\infty \left.\log \ell(y|\mu(x, \theta))\right|_{\theta=\theta(x,y)}f_Y(y) \le \sum_{y=0}^\infty \left.\log \ell(y|\mu)\right|_{\mu=\mu(y)}f_Y(y) \le M_Y.
\]
And we assume the distribution of $X$ satisfies the regularity condition, as well:
\[
\int_x\log\ell(x)f_X(x)dx \le M_X.
\]
We can therefore conclude that
\[
    \int_x \sum_{y=0}^\infty \left.\log \ell(x, y|\theta)\right|_{\theta=\theta(x,y)}f_Y(y) f_X(x)dx
    \le M_X + \int_x M_Y f_X(x)dx
    =M_X + M_Y.
\]
Since the preceding argument applies to both $\theta_{\max}$ and $\theta_{\min}$, we have
\[
\mathbb{E}_{\theta_0}[d(X, Y)] < \infty.
\]

{\bf Condition (5)}. 
% take the neighborhood to be all of $\Theta$. The second part of the condition, that $\ell(x,y\vert\theta) > 0$, is true by inspection for bounded $\theta$.
% For the first part of the condition, we begin by noting that we can write the Hessian of the likelihood function in terms of the log likelihood function:
% \begin{align}
%     \nabla^2_\mathbf{\theta} \ell_\mathbf{\theta} (Y,X)
%     &= \nabla^2_\mathbf{\theta} \ell_\mathbf{\theta} (Y\vert X) \cdot \ell (X)\\
%     &= \ell(X) \nabla^2_\mathbf{\theta} \exp(\log \ell_\mathbf{\theta} (Y\vert X))\\
%     &= \ell(X) \cdot  \ell_\mathbf{\theta} (Y\vert X) \left( \nabla_\mathbf{\theta}^2 \log \ell_\mathbf{\theta} (Y\vert X) 
%     + \nabla_\mathbf{\theta} \log \ell_\mathbf{\theta} (Y\vert X)\left(\nabla_\mathbf{\theta} \log \ell_\mathbf{\theta} (Y\vert X)\right)^\top
%     \right)
% \end{align}
% where we have also used the fact that $X$ does not depend on $\theta$. This expression is the sum and product of continuous functions: the likelihood $\ell(X)$ is constant in $\mathbf{\theta}$, the likelihood $\ell_\mathbf{\theta}(Y\vert X)$ is continuous in $\mathbf{\theta}$, the first derivative of $\log \ell_\mathbf{\theta}(Y\vert X)$ is differentiable (as evidenced by the fact that we computed its derivative in the proof of Theorem \ref{thm:fullGenerality}), and finally the second derivative of the log likelihood is continuous by inspection.
% \textcolor{TealBlue}{Peng: Thanks Jennifer! I think this works, but I think we could also simply argue by the compositional rule.
Using the expression \eqref{eq:likelihood_demp} and the fact that $\ell(y|\mu)$ is twice differentiable with respect to $\mu$, and $\mu$ is twice differentiable with respect to $\theta$, we know that $\ell(x, y|\theta)$ is twice differentiable with respect to $\theta$.
Moreover, since $\ell(y|\mu)$ is always positive in the neighborhood of $\mu$ and $\ell(x)$ is always positive, we know that $\ell(x, y | \theta)$
is always positive in the neighborhood of $\theta$.

To verify {\bf Condition (6)}, we begin with the decomposition of the likelihood and then use the fact that $\mu(x,\theta)=p(x, \theta)\lambda(x, \theta)$ describes and \eqref{eq:likelihood_demp} to %\sandy{"describes" needs an object, it needs to be removed, or something is wrong in the preceding sentence. Please correct.}
write
\[
\frac{\partial \ell(x, y | \theta)}{\partial \theta_i} = \ell(x)\frac{\partial \ell(y | \mu(x, \theta))}{\partial\theta_i} = \ell(x)\frac{\partial \ell(y | \mu)}{\partial\mu}\frac{\partial\mu(x, \theta)}{\partial\theta_i}.
\]
Since $\Theta$ is a compact set, the upper and lower bounds of the partial derivative can be attained. Here, we use same notions $\theta(x, y)$, $\theta_{\max}(x,y)$ and $\theta_{\min}(x,y)$ as in the proof of condition (4). Our dominating function can now be written as
\[
e(x, y) = \max\left\{\left.\ell(x)\frac{\partial \ell(y | \mu)}{\partial\mu}\frac{\partial\mu(x, \theta)}{\partial\theta_i}\right|_{\theta=\theta_{\max}(x, y)},
-\left.\ell(x)\frac{\partial \ell(y | \mu)}{\partial\mu}\frac{\partial\mu(x, \theta)}{\partial\theta_i}\right|_{\theta=\theta_{\min}(x, y)}\right\}.
\]
Since $\nu(x,y)$ is the uniform measure over $x$ and $y$, we have
\[
\int \left.\frac{\partial \ell(x, y | \theta)}{\partial \theta_i} \right|_{\theta=\theta(x,y)}d\nu(x, y) =\int_x \left.\ell(x)\frac{\partial\mu(x, \theta)}{\partial \theta_i}\sum_{y=0}^\infty \frac{\partial\ell(y|\mu)}{\partial \mu}\right|_{\theta=\theta(x, y)} dx.
\]
The Poisson likelihood satisfies the regularity condition because it is a generalized linear model, and therefore we have
\[
\left.\sum_{y=0}^\infty\frac{\partial\ell(y|\mu)}{\partial \mu}\right|_{\theta=\theta(x, y)} \le \left.\sum_{y=0}^\infty\frac{\partial\ell(y|\mu)}{\partial \mu}\right|_{\mu=\mu(y)} \le
M
\]
We can now write an upper bound on the quantity of interest:
\[\begin{aligned}
\int\left.\frac{\partial \ell(x, y | \theta)}{\partial \theta_i} \right|_{\theta=\theta(x,y)}d\nu(x, y) &\le M\int_x \left.\ell(x)\frac{\partial\mu(x, \theta)}{\partial \theta_i} \right|_{\theta=\theta(x, y)} dx.
\end{aligned}\]
Since $\mu = p\cdot\lambda$ is the product of an exponential and an expit 
function, the partial derivative of $\mu$ with respect to $\theta$ can grow at most as fast as exponential function, and it will be dominated by the density function of Gaussian distribution $\ell(x)$. Therefore, we conclude that
\[
\int e(x, y) d\nu(x, y) < +\infty,
\]
which shows that Condition (6) is satisfied.

{\bf Condition (7)} is satisfied by inspection of the Fisher Information Matrix, which is computed in the proof of Theorem \ref{thm:fullGenerality}.

The proof of {\bf Condition (8)} and {\bf Condition (9)}
% To show that  is satisfied, we use the upper bound on the Fisher Information Matrix provided in Theorem \ref{thm:fullGenerality} \textcolor{red}{Finish}
follows from  the same arguments used to prove {\bf Condition (4)} 
and {\bf Condition (6)}, respectively. 

Since all conditions are satisfied, the conclusion follows immediately. 
% We see that all of the conditions are satisfied, and conclude that the MLE is asymptotically distributed according to 
% \begin{align}
%     \mathbf{\hat\theta}_{MLE} \sim \mathcal{N}\left(\mathbf{\theta}, \mathcal{I}(\mathbf{\theta})^{-1}\right) 
% \end{align}

\section{Proof of Theorem~\ref{thm:fullGenerality}}
We will prove this result using the Cram\'er-Rao lower bound. The result states that under certain regularity conditions (which we show are satisfied in Lemma \ref{lemma:mleNormConditions}), 
\begin{align}
    \text{Cov}(\bm{\hat\theta}) \succeq \frac{1}{n}
    \mathcal{I}(\bm\theta)^{-1},\label{eqn:cramerRao},
\end{align}
where $\mathcal{I}(\theta)$ is the Fisher Information matrix, defined as
\begin{align}
    \mathcal{I}(\bm\theta)
    &:=
    \E_{\bm\theta} \left[ \nabla_{\bm\theta} \log \ell_{\bm\theta}(Y, X) \nabla_{\bm\theta} \log \ell_{\bm\theta}(Y, X)^\top\right],
\end{align}
where $\ell_{\bm\theta}(Y, X)$ is the likelihood of the observed data under the parameters $\theta$. Under mild regularity conditions, which we show in Lemma \ref{lemma:mleNormConditions} are satisfied in this instance, we can write the Fisher information matrix as the negative Hessian of the log likelihood function
\begin{align}
    \mathcal{I}(\bm\theta)
    &=
    - \E_{\bm\theta} \left[ \nabla_{\bm\theta}^2 \log \ell_{\bm\theta}(Y, X)\right].
\end{align}
We compute the covariance in $\bm{\hat\theta}$ with respect to the randomness in both $Y$ and $X$:
\begin{align}
    \text{Cov}(\bm{\hat\theta})
    &\geq \frac{1}{n}  \mathcal{I}^{-1} (\bm{\theta}) \\
    &= \frac{1}{n}\E_{X,Y} \left[ -\nabla_{\bm\theta}^2\log \ell_{\bm\theta}(Y,X) \right]^{-1}\\
    &= \frac{1}{n}\E_{X,Y} \left[ -\nabla_{\bm\theta}^2 \left(\log \ell_{\bm\theta}(X) + \log\ell_{\bm\theta}(Y|X)\right) \right]^{-1}\\
    &= \frac{1}{n}\E_{X,Y} \left[-\nabla_{\bm\theta}^2 \log\ell_{\bm\theta}(Y|X) \right]^{-1},
\end{align}
where we have used the fact that the distribution of $X$ is independent of the parameters $\bm\theta$. Next, we compute the Hessian of the negative log conditional likelihood. The log conditional likelihood under this model is given by
\begin{align}
    \log\ell_{\bm\theta}(Y|X)
    &= -\log(Y!) - e^{X_{\lambda}\theta_\lambda}\left( \frac{\exp(X_p \theta_p)}{1+\exp(X_p \theta_p)}\right) + Y\left(X_\lambda \theta_\lambda + \log\left(\frac{\exp(X_p \theta_p)}{1+\exp(X_p \theta_p)}\right)\right),
\end{align}
and the Hessian of the negative log conditional likelihood has components
\begin{align}
-\nabla_{\bm\theta}^2 \log\ell_{\bm\theta}(Y|X) 
&= \begin{bmatrix}
-\frac{\partial^2}{\partial^2\theta_\lambda} \log\ell_{\bm\theta}(Y|X) & 
-\frac{\partial^2}{\partial\theta_\lambda\partial\theta_p} \log\ell_{\bm\theta}(Y|X)\\
-\frac{\partial^2}{\partial\theta_p\partial\theta_\lambda} \log\ell_{\bm\theta}(Y|X) &
-\frac{\partial^2}{\partial^2\theta_p} \log\ell_{\bm\theta}(Y|X)
\end{bmatrix}.\label{eqn:negHessian}.
\end{align}
We will bound each of these quantities independently. Note that the matrix is symmetric; we begin by showing that the off-diagonal entry is zero:
\begin{align}
    \mathbb{E}\left[ -\frac{\partial^2}{\partial\theta_\lambda\partial\theta_p} \log\ell_{\bm\theta}(Y|X) \right]
    &=\mathbb{E}\left[ e^{X_\lambda\theta_\lambda}\frac{\exp(X_p \theta_p)}{(1+\exp(X_p \theta_p))^2} X_\lambda X_p \right]\\
    &=\mathbb{E}\left[ e^{X_\lambda\theta_\lambda} X_\lambda\right]
    \mathbb{E}\left[\frac{\exp(X_p \theta_p)}{(1+\exp(X_p \theta_p))^2} X_p \right].
\end{align}
Note that the second expectation is over an odd function. Since we assumed $X_p$ is symmetric around zero, this term is zero, and 
\begin{align}
    \mathbb{E}\left[ -\frac{\partial^2}{\partial\theta_\lambda\partial\theta_p} \log\ell_{\bm\theta}(Y|X) \right] &= 0.
\end{align}

Next, we compute the first diagonal entry in the matrix. We begin by separating it into terms that depend on $X_\lambda$ and terms that depend on $X_p$:
\begin{align}
    \E\left[ -\frac{\partial^2}{\partial^2\theta_\lambda} \log\ell_{\bm\theta}(Y|X)\right]
    &= \mathbb{E}\left[\sum_i e^{X_\lambda\theta_\lambda}\frac{\exp(X_p \theta_p)}{1+\exp(X_p \theta_p)} X_\lambda^2\right]\\
    &= \mathbb{E} \left[ e^{X_\lambda \theta_\lambda} X_\lambda^2 \right] \mathbb{E}\left[ \frac{e^{X_p \theta_p}}{1 + e^{X_p \theta_p}} \right]\\
    &= \mathbb{E} \left[ e^{X_\lambda \theta_\lambda} X_\lambda^2 \right] \mathbb{E}[p].
\end{align}
Next, we evaluate the first expectation using the known distribution of $X_\lambda$ and then completing the square:
\begin{align}
\E\left[ -\frac{\partial^2}{\partial^2\theta_\lambda} \log\ell_{\bm\theta}(Y|X)\right]
    &= \mathbb{E}[p]
    \int_{-\infty}^\infty \frac{1}{\sqrt{2\pi\sigma_\lambda^2}} x^2 \exp\left(-\frac{(x-\mu_\lambda)^2}{2\sigma_\lambda^2} + x\theta_\lambda\right)dx\\
    &= \mathbb{E}[p]\int_{-\infty}^\infty \frac{1}{\sqrt{2\pi\sigma_\lambda^2}} x^2 e^{\mu_\lambda \theta_\lambda + \sigma_\lambda^2 \theta_\lambda^2/2} \exp\left( -\frac{1}{2\sigma_\lambda^2} (x - (\mu_\lambda + \sigma_\lambda^2\theta_\lambda))^2 \right) dx\\
    &= \mathbb{E}[p]e^{\mu \theta_\lambda + \sigma_\lambda^2 \theta_\lambda^2/2} \mathbb{E}_{x\sim\mathcal{N}(\mu_\lambda + \sigma_\lambda^2 \theta_\lambda, \sigma_\lambda^2)}[x^2]\\
    &=  \mathbb{E}[p]\E\left[\lambda\right]\left( (\mu_\lambda + \sigma_\lambda^2 \theta_\lambda)^2 + \sigma_\lambda^2 \right).
\end{align}
Now, all that remains is to upper bound the final term in the Hessian of the negative log likelihood:
\begin{align}
    \E\left[ -\frac{\partial^2}{\partial^2\theta_p} \log\ell_{\bm\theta}(Y|X)\right]
    &= \E\left[\frac{\exp(X_p\theta_p)}{(1+\exp(X_p\theta_p))^3}\left(e^{X_\lambda\theta_\lambda} \left(1-e^{X_p\theta_p}\right) + Y\left(1+e^{X_p\theta_p}\right)\right) X_p X_p\right].
\end{align}
We begin with the tower rule of expectation, using the fact that $\E[Y|X] = \frac{\exp(X_p \theta_p)}{1+\exp(X_p \theta_p)}e^{X_\lambda \theta_\lambda}$.
\begin{align}
    \E\left[ -\frac{\partial^2}{\partial^2\theta_p} \log\ell_{\bm\theta}(Y|X)\right]
    &= \E\left[\E\left[\frac{\exp(X_p\theta_p)}{(1+\exp(X_p\theta_p))^3}\left(e^{X_\lambda\theta_\lambda} \left(1-e^{X_p\theta_p}\right) + Y\left(1+e^{X_p\theta_p}\right)\right) X_p X_p \Big| X\right]\right]\\
    &= \E\left[\frac{\exp(X_p\theta_p)}{(1+\exp(X_p\theta_p))^3}\left(e^{X_\lambda\theta_\lambda} \left(1-e^{X_p\theta_p}\right) + \E\left[Y| X\right]\left(1+e^{X_p\theta_p}\right)\right) X_p^2 \right]\\
    &= \E\left[e^{X_\lambda \theta_\lambda}\frac{\exp(X_p\theta_p)}{(1+\exp(X_p\theta_p))^3} X_p^2 \right]
\end{align}
Next, we use the fact that $X_p$ and $X_\lambda$ are independent to separate the expectation over $X_p$ from the expectation over $X_\lambda$:
\begin{align}
    \E\left[ -\frac{\partial^2}{\partial^2\theta_p} \log\ell_{\bm\theta}(Y|X)\right]
    &= \E\left[e^{X_\lambda \theta_\lambda}\right] \E\left[\frac{\exp(X_p\theta_p)}{(1+\exp(X_p\theta_p))^3} X_p^2 \right]\\
    &= \E[\lambda] \E\left[\frac{\exp(X_p\theta_p)}{(1+\exp(X_p\theta_p))^3} X_p^2 \right].
\end{align}

We will now show that, \textit{regardless of the distribution of $X_p$}, this remaining expectation is less than $\frac{1}{2\theta_p^2}$. Our strategy will be to upper bound the expectation by the maximum value of its argument. We have
\begin{align}
    \mathbb{E} \left[\frac{\exp(X_p\theta_p)}{(1+\exp(X_p\theta_p))^3} X_p^2  \right]
    &\leq \max_x \frac{\exp(x \theta_p)}{(1+\exp(x \theta_p))^3} x^2 \\
    &= \frac{1}{\theta_p^2}\max_x \frac{\exp(x \theta_p)}{(1+\exp(x \theta_p))^3} (x\theta_p)^2.
\end{align}
Now, let $u=x\theta_p$:
\begin{align}
    \mathbb{E} \left[\frac{\exp(X_p\theta_p)}{(1+\exp(X_p\theta_p))^3} X_p^2  \right]
    &\leq \frac{1}{\theta_p^2}\max_u \frac{\exp(u)}{(1+\exp(u))^3} u^2.
\end{align}
Here, we provide a simple upper bound for quantity
\[
C \triangleq \max_{u} \frac{\exp(u)}{(1 + \exp(u))^3}u^2.
\]
We divide the maximization problem over cases when $u$ is positive and when $u$ is non-positive. When $u > 0$, we know that
\[
\frac{\exp(u)}{(1 + \exp(u))^3}u^2 = \frac{\exp(u)}{(1 + \exp(u))^2}\frac{u^2}{1 + \exp(u)} \le
\frac{\exp(u)}{(1 + \exp(u))^2} \le \frac{1}{4}.
\]
And when $u \le 0$, we have
\[\begin{aligned}
\frac{\exp(u)}{(1 + \exp(u))^3}u^2 &=
\frac{1}{1 + \exp(u)}\frac{u^2}{(1 + \exp(u))(1 + \exp(-u))}\\
&= \frac{1}{1 + \exp(u)}\frac{u^2}{2 + \exp(u) + \exp(-u)}\\
&\le \frac{u^2}{2 + \exp(u) + \exp(-u)}\\
&\le \frac{u^2}{4 + u^2 + \frac{u^4}{12}}
\le \frac{\sqrt{3}}{\sqrt{3} + 2}\le \frac{1}{2}.
\end{aligned}\]
Therefore, we know that $C \le 1/2$.

% The $u$ that maximizes this function satisfies
% \begin{align}
%     0 &= \frac{d}{du} \frac{\exp(u)}{(1+\exp(u))^3} u^2 \\
%     &= -\frac{ue^{u}(-u + 2e^{u}(u - 1)-2)}{(1+e^{u})^4}\\
%     0 &= u(-u + 2e^{u}(u - 1)-2)
% \end{align}
% We note that $u=0$ is a local minimum of this function (at which point the function value is 0; as is also true at $u=\pm\infty$), so we divide by $u$ and continue,
% \begin{align}
%     0 &= -u + 2e^{u}(u - 1)-2
% \end{align}
% This equation has two roots, $u_l\in[-2.6, -2.5]$ and $u_r\in [1.4, 1.5]$. We can rewrite the maximum over $u$ as the maximum over this restricted set,
% \begin{align}
%     \mathbb{E}\left[\frac{\exp(X_p\theta_p)}{(1+\exp(X_p\theta_p))^3} X_p^2  \right]
%     &\leq \frac{1}{\theta_p^2} \max\left\{\max_{u\in[-2.6, -2.5]} u^2\frac{e^u}{(1+e^u)^4}, \max_{u\in[1.4, 1.5]} u^2\frac{e^u}{(1+e^u)^4} \right\}\\
%     &\leq \frac{1}{\theta_p^2} \max\left\{\max_{u\in[-2.6, -2.5]} u^2\cdot \max_{u\in[-2.6, -2.5]}\frac{e^u}{(1+e^u)^4}, \max_{u\in[1.4, 1.5]} u^2 \cdot \max_{u\in[1.4, 1.5]}\frac{e^u}{(1+e^u)^4} \right\}\\
%     &= \frac{1}{\theta_p^2} \max\{0.5, 0.08\}\\
%     &= \frac{1}{2\theta_p^2}
% \end{align}
% We conclude that, regardless of the distribution of $X_p$, we have
% \begin{align}
%     \mathbb{E}\left[ -\frac{\partial^2}{\partial^2\theta_p} \log\ell_{\bm\theta}(Y|X)\right] &\leq
%     0.5 n e^{\mu_\lambda \theta_\lambda + \sigma_\lambda^2 \lambda^2 / 2}
%     \theta_p^{-2}.
% \end{align}
We conclude that, regardless of the distribution of $X_p$, we have
\begin{align}
    \mathbb{E}\left[ -\frac{\partial^2}{\partial^2\theta_p} \log\ell_{\bm\theta}(Y|X)\right] &\leq
    \tfrac{1}{2} \E[\lambda]
    \theta_p^{-2}.
\end{align}
We have shown that the Fisher information matrix is diagonal and lower bounded its diagonal entries, so the conclusion of the theorem follows from Eqn \eqref{eqn:cramerRao}.

\end{document}